\documentclass[a4paper,11pt]{article}

\usepackage{jheppub} 
\usepackage[normalem]{ulem}  
\usepackage{amsmath}
\usepackage{comment}

\newcommand{\cG}{\mathcal{G}}

\newcommand{\cD}{\mathcal{D}}
\newcommand{\bsx}{\boldsymbol{x}}

\preprint{KOBE-COSMO-24-06, UT-Komaba/24-11}

\title{
Gravitational EFT for dissipative open systems
}

\author[a,b,c]{Pak Hang Chris Lau,}

\emailAdd{phcl2@gbu.edu.cn}

\affiliation[a]{School of Sciences, Great Bay University, Dongguan 523000, China}
\affiliation[b]{Department of Physics, Osaka University, Toyonaka, Osaka 560-0043, Japan.}
\affiliation[c]{Department of Physics, Kobe University, Kobe 657-8501, Japan}
\affiliation[d]{Graduate School of Arts and Sciences, University of Tokyo, Komaba, Meguro-ku, Tokyo 153-8902, Japan}

\author[c,d]{Kanji Nishii}
\emailAdd{kanji.nishii@stu.kobe-u.ac.jp}

\author[d]{and Toshifumi Noumi}
\emailAdd{tnoumi@g.ecc.u-tokyo.ac.jp}

\abstract{
We elaborate on the effective field theory (EFT) construction for dissipative open systems coupled to dynamical gravity, in light of recent developments on the EFT of dissipative hydrodynamics (HydroEFT). Our construction is based on the Schwinger-Keldysh formalism and its symmetries as well as microscopic unitarity.  A key aspect of dynamical gravity is that gravity couples to all degrees of freedom universally, hence the EFT has to take into account the energy-momentum tensor of the environment to which the energy escapes from the dissipative system of interest. We incorporate this effect by modeling the environment based on HydroEFT, assuming validity of the derivative expansion of the environment sector. For illustration, we apply our EFT recipe to a dissipative scalar field coupled to dynamical gravity that can be used, e.g., for dissipative inflation. In particular we quantify impacts of fluctuations in the environment sector on the scalar dynamics. We also apply the same framework to dissipative gravity, discussing dissipative gravitational waves and the generalized second law of black hole thermodynamics.
}

\begin{document} 
\setcounter{tocdepth}{2}
\maketitle
\flushbottom

\newpage

\section{Introduction}
The Schwinger-Keldysh (SK) formalism is widely employed as a basic tool for analyzing non-equilibrium systems. This framework can be applied to describe the real-time evolution of finite-temperature systems and quantum open systems. It has also been acknowledged as a systematic approach for constructing effective field theories (EFTs) that explicitly incorporate fluctuations and dissipation. Recent advancements in non-equilibrium systems based on the SK formalism have been remarkable~\cite{Crossley:2015evo,Glorioso:2016gsa,Glorioso:2017fpd,Liu:2018kfw,Sieberer:2015hba,Haehl:2015pja,Jensen:2017kzi,Jensen:2018hhx,Minami:2015uzo,Hongo:2018ant,Hayata:2018qgt,Hongo:2019qhi,Landry:2019iel,Landry:2020ire,Landry:2020tbh,Landry:2021kko,Fujii:2021nwp,Polonyi:2019kjn,Ota:2024mps,Salcedo:2024smn,Hongo:2024brb,Salcedo:2024nex,Akyuz:2023lsm,Sieberer:2015svu,Chen-Lin:2018kfl,Chao:2020kcf}. For instance, active research has been conducted on the EFT of hydrodynamics (HydroEFT) based on the dynamical KMS symmetry that characterizes local equilibrium in~\cite{Crossley:2015evo,Glorioso:2016gsa,Glorioso:2017fpd,Liu:2018kfw,Sieberer:2015hba,Haehl:2015pja,Jensen:2017kzi,Jensen:2018hhx}.
Spontaneous symmetry breaking in open systems and EFTs thereof are studied in~\cite{Minami:2015uzo,Hongo:2018ant,Hayata:2018qgt,Hongo:2019qhi,Landry:2019iel,Landry:2020ire,Landry:2020tbh,Landry:2021kko,Fujii:2021nwp}. More recently, following these developments, application to the open system type EFT of cosmic inflation, the open system EFT of inflation, has also been initiated~\cite{Hongo:2018ant,Salcedo:2024smn} along the line of the EFT of inflation~\cite{Cheung:2007st}\footnote{See~\cite{LopezNacir:2011kk} for a pioneering work on dissipation in EFT of inflation that does not use the SK formalism.}, even though current studies are limited to EFTs on a fixed background spacetime without dynamical gravity.

\medskip
The purpose of this paper is to elaborate on the EFT construction for dissipative open systems coupled to dynamical gravity. This is a natural extension of the aforementioned developments and necessary when exploring their application to gravitational systems, such as those relevant to cosmology and black hole physics. A key aspect of
dynamical gravity is that it couples to all degrees of freedom universally. Typically, open system EFTs are described only by the system variables of interest and in particular their energy-momentum tensor is not conserved due to the open system nature. On the other hand, to incorporate dynamical gravity, we need to take into account the energy-momentum tensor of the environment sector appropriately: The energy-momentum tensor appearing in the Einstein equation is that for the total system which enjoys the conservation law. At the same time, introduction of the energy-momentum tensor of the environment sector accounts for the energy outflow caused by dissipation, which is necessary to follow the gravitational dynamics. Technically, this simple fact turns out to be well captured by the doubled diffeomorphism symmetries of the SK formalism.

\medskip
In the SK formalism, the closed-time-path is used to describe the time-evolution of the system (see Fig.~\ref{fig1}), resulting in the doubling of the action in the generating functional to account for the forward and backward time-evolutions. Consequently, each component of the doubled action for dynamical gravity has its own diffeomorphism symmetry. These symmetries can be decomposed into diagonal and off-diagonal parts within the EFT framework, which are called the physical and noise diffeomorphisms, respectively~\cite{Crossley:2015evo,Glorioso:2017fpd}.
Among them, the noise diffeomorphism symmetry is responsible for conservation of the energy-momentum tensor and indeed it is broken by the EFTs of dissipative open systems, as long as we focus only on the system sector.
On the other hand, EFTs with dynamical gravity have to respect both of the physical and noise diffeomorphism symmetries, as we will explain in Sec.~\ref{sec:diff_sym}. It then turns out to be convenient to introduce St\"{u}ckelberg fields that non-linearly realize the noise diffeomorphism symmetry broken by dissipation. In Sec.~\ref{sec:diss_sca_grav}, we demonstrate that we need to promote these St\"{u}ckelberg fields to dynamical fields and introduce the energy-momentum tensor of the environment in order to describe dissipation in the presence of dynamical gravity. This is the technical side of the above story.

\medskip
While introduction of the energy-momentum tensor of the environment is a general requirement of dynamical gravity, modeling of the environment is necessary for completing the EFT construction. Since St\"{u}ckelberg fields for the noise diffeomorphism symmetry are required to be dynamical, it is natural to introduce those for the physical diffeomorphism to form a pair of doubled SK path-integral variables. Notably, HydroEFT uses these St\"{u}ckelberg fields as slow-variables and offers a simple and natural model for the environment, which we utilize when constructing the EFT and discussing its property concretely.

\medskip
The above considerations are summarized by the following recipe of the SKEFT for dissipative open sytems coupled to dynamical gravity:
\begin{enumerate}
\item Construct a SK effective action of the target dissipative open system in a physical diffeomorphism invariant manner. In particular, dissipation terms are captured by effective operators that break the noise diffeomorphism symmetry.

\item Introduce the St\"{u}ckelberg fields to restore the noise diffeomorphism symmetry. Note that the St\"{u}ckelberg fields introduced here play a role of auxiliary fields that prohibit dissipation, unless the environment sector is introduced.

\item Make the St\"{u}ckelberg fields dynamical by introducing the energy-momentum tensor of the environment sector and turn on dynamical gravity.

\item The EFT constructed in this manner can be used to describe fluctuations in the system sector, the environment sector, and the gravity sector. In general the three sectors are coupled with each other.

\item One may also use the EFT to quantify the conditions under which fluctuations in the environment are negligible. The EFT after decoupling the environment is formulated in terms of the system variables (and gravity) alone.

\end{enumerate}
In this paper we perform this EFT construction concretely for a dissipative scalar coupled to dynamical gravity as well as for dissipative gravity with some explicit applications.

\medskip
This paper is organized as follows: In Sec.~\ref{sec:SK_review}, we begin with a review of the general framework of the SK formalism, using a toy oscillator model for illustration. In Sec.~\ref{sec:diff_sym}, we then provide a detailed discussion of the diffeomorphism symmetries within the closed-time-path contour. In Sec.~\ref{sec:diss_sca_grav}, we elaborate on the above recipe of the SKEFT for dissipative systems coupled to dynamical gravity using a scalar matter field as an illustrative example. The same framework is applied to dissipative gravity  in Sec.~\ref{sec:diss_grav}.
As an application, we discuss dissipative gravitational waves and the generalized second law of black hole thermodynamics.

\section{Schwinger-Keldysh formalism: a pedagogical review}
\label{sec:SK_review}

We begin by a review of the Schwinger-Keldysh (SK) formalism (also known as the closed-time-path formalism) and its application to dissipative systems, using a simple model of harmonic oscillators. The aim of this section is to share the necessary physical intuitions and background technical information of the present paper in a rather compact and self-contained manner. For more details of the SK formalism, we refer the readers to, for example, Refs.~\cite{Kamenev,BenTov:2021jsf,Liu:2018kfw}.
Readers familiar with the SK formalism, especially coarse graining in this framework, can safely skip Secs.~\ref{subsec:micro}-\ref{subsec:consistency} and go to Sec.~\ref{subsec:general}, where general properties of the SK effective field theory (SKEFT) are summarized.

\subsection{Microscopic description}
\label{subsec:micro}

\begin{figure}[t] 
	\centering 
	\includegraphics[width=9.5cm]{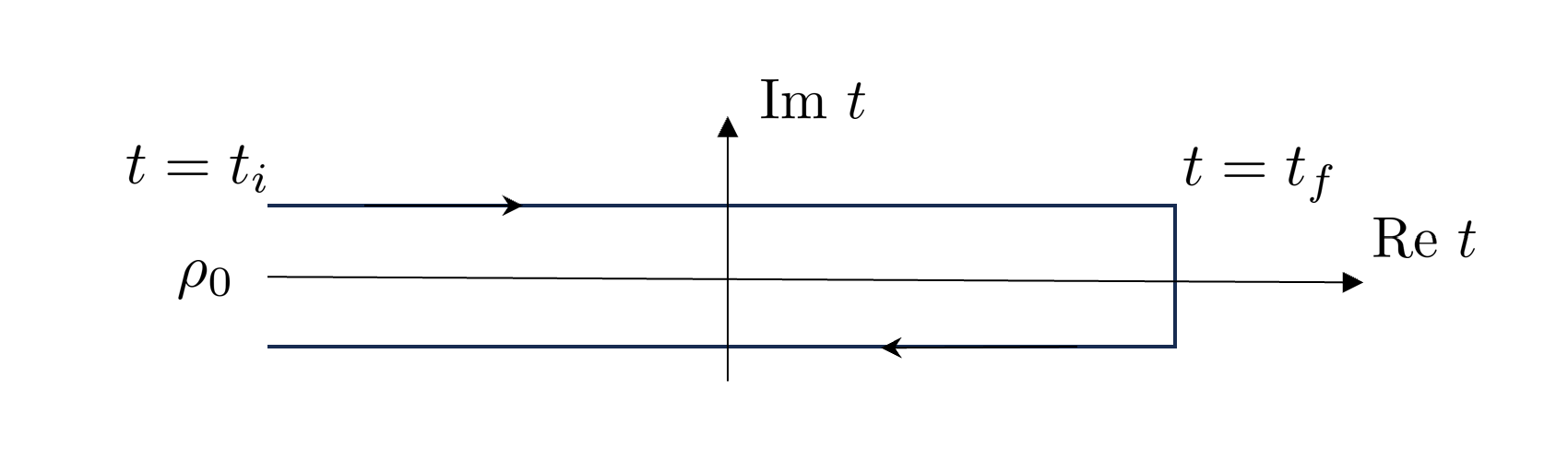} 
	\caption{The closed-time-path on which the generating functional is defined.}
\label{fig1}
\end{figure}

The SK formalism offers a framework for evaluating correlation functions in non-equilibrium systems. A basic quantity is the closed-time-path generating functional $W[J_1,J_2]$ defined in the operator formalism as
\begin{align}
\label{W_op}
e^{iW[J_1,J_2]}&={\rm Tr}\left[
\rho_0\,
U_{J_2}^\dagger(t_f,t_i)\,
U_{J_1}(t_f,t_i)
\right]
\,,
\end{align}
where $U_J(t_2,t_1)$ is the time-evolution operator from $t_1$ to $t_2$ in the presence of external sources collectively denoted by $J$. $\rho_0$ is the initial density operator inserted at $t_1$. In the path-integral form, $W[J_1,J_2]$ is given by
\begin{align}
\label{generating_function_PI}
e^{iW[J_1,J_2]}&=\int \cD\Phi_1\cD\Phi_2 \,e^{i(S[\Phi_1;J_1]-S[\Phi_2;J_2])}\,,
\end{align}
where $S[\Phi;J]$ is the microscopic action of dynamical degrees of freedom, collectively denoted by $\Phi$. The SK formalism uses a pair of path-integral variables, $\{\Phi_1,\Phi_2\}$ for each dynamical degree of freedom 
$\Phi$, with each variable assigned to a (horizontal) section of the contour in Fig.~\ref{fig1}. As a result, the number of degrees of freedom is doubled compared to the in-out formalism. The boundary conditions at the initial time $t_i$ are specified by the initial density operator $\rho_0$. On the other hand, the boundary conditions at the final time $t_f$ are $\Phi_1(t_f)=\Phi_2(t_f)$.

\paragraph{A toy model.}

Many aspects of the SKEFT can be captured by a simple model of a harmonic oscillator $X$ coupled to an environment modeled by a set of harmonic oscillators $Y_n$ ($n=1,\ldots,N$) with the following microscopic action\footnote{
This model is often used as an introduction to the SKEFT. See, e.g., Refs.~\cite{Kamenev,Hongo:2018ant} for more details.}:
\begin{align}
S[X,Y_n;J]&=\int dt\bigg[\,
\left(\frac{1}{2}\dot{X}^2-\frac{1}{2}\omega_0^2X^2\right)+\sum_{n=1}^N\left(\frac{1}{2}\dot{Y}_n^2-\frac{1}{2}\omega_nY_n^2\right)+\sum_{n=1}^Ng_nXY_n
\nonumber
\\
&\qquad\qquad
+(\text{source terms})\bigg]
\,, \label{eqn:osc_micro}
\end{align}
where the overhead dot represents the time derivative $\dot{f} \ \equiv\frac{d }{dt}f$, $\omega_n$ ($n=0,\ldots,N$) are the frequencies of the oscillators and $g_n$ are the couplings between the system oscillator $X$ and the set of environment oscillators $Y_n$. Details of the source terms denoted by the last term are not important in our discussion and are kept implicit.

\medskip
The corresponding path-integral description is
\begin{align}
\label{generating_function_PI_grav}
e^{iW[J_1,J_2]}&=\int \cD X_1\cD X_2\cD Y_{n1}\cD Y_{n2}
\,e^{iS_{\rm micro}[X_1,Y_{n1},X_2,Y_{n2};J_1,J_2]} \,,
\end{align}
with appropriate boundary conditions at $t=t_i$ and $t=t_f$. Here we have introduced a pair of path-integral variables $\{(X_1,Y_{n1}),(X_2,Y_{n2})\}$ for the set of dynamical variables $\{(X,Y_n)\}$ appearing in \eqref{eqn:osc_micro}. Also, the path-integral weight is
\begin{align}
\label{micro_oscillators}
S_{\rm micro}[X_1,Y_{n1},X_2,Y_{n2};J_1,J_2]
=S[X_1,Y_{n1};J_1]-S[X_2,Y_{n2};J_2]\,.
\end{align}
We emphasize that the microscopic SK action $S_{\rm micro}$ is factorized into the $(X_1,Y_{n1};J_1)$ part and the $(X_2,Y_{n2};J_2)$ part, whereas the boundary conditions mix the two sectors.

\subsection{Green's functions of the environment oscillators}

Next we would like to write down the effective theory of the system oscillator $X$. In our illustrative example, where only terms up to quadratic order are present, it is straightforward to integrate out the environment oscillators $Y_{n1}$ and $Y_{n2}$ to obtain the following effective action in the Fourier space:
\begin{align}
\label{EFT_XY}
iS_{\rm eff}=-\frac{1}{2}\sum_{s,s'=1,2}\int\frac{d\omega}{2\pi}
\left[
X_s(-\omega)D_{ss'}(\omega;\omega_0^2)X_{s'}(\omega)
+\sum_{n=1}^N g_n^2X_s(-\omega)G_{ss'}(\omega;\omega_n^2)X_{s'}(\omega)
\right]\,,
\end{align}
where we have suppressed the source terms for visual clarity. $D_{ss'}(\omega;\Omega^2)$ is the $2\times 2$ kinetic matrix of the harmonic oscillator with frequency $\Omega$ and $G_{ss'}(\omega;\Omega^2)$ are the Green's functions given by the inverse matrix of $D_{ss'}(\omega;\Omega^2)$. 

\paragraph{Off-shell Green's functions.}

To elaborate on the effects of the environment, let us take a closer look at the Green's functions. First, the kinetic matrix $D_{ss'}(\omega;\omega_n)$ is of the form,
\begin{align}
D_{ss'}(\omega;\omega_n^2)=\left(\begin{array}{cc}-i(\omega^2-\omega_n^2) & 0 \\0 & i(\omega^2-\omega_n^2)\end{array}\right)\,.
\end{align}
For now, we will neglect the 
$i\epsilon$-prescription typically required to address the on-shell dynamics at $\omega=\omega_0$. The corresponding Green's functions $G_{ss'}(\omega)$ are
\begin{align}
G_{ss'}(\omega;\omega_n^2)=\left(D_{ss'}(\omega;\omega_n^2)\right)^{-1}=\left(\begin{array}{cc}\displaystyle\frac{i}{\omega^2-\omega_n^2} & 0 \\0 &\displaystyle -\frac{i}{\omega^2-\omega_n^2}\end{array}\right)
\quad
{\rm for}
\quad
\omega\neq\omega_n\,.
\end{align}
Note that these off-shell Green's functions are diagonal due to the diagonal nature of the kinetic matrix $D_{ss'}$, so that the $X_1$ sector and the $X_2$ sector are factorizable even in the effective action, as long as there are no on-shell contributions of the environment.

\medskip
In the field theory language, this means that the effective theory has no mixing between the two sets of path-integral variables as long as on-shell particle production in the corse-grained environment sector does not occur. In this situation, one can simply use the in-out effective action to compute correlation functions in the system sector. For example, the EFT of inflation~\cite{Cheung:2007st,Weinberg:2008hq} captures effective interactions sourced by heavy fields whose particle production during inflation is exponentially suppressed by the Boltzmann factor.

\paragraph{Full Green's functions.}

On the other hand, to derive the full Green's functions including the on-shell component, we need to identify the initial conditions and also be careful about the $i\epsilon$-prescription. For example, when the initial density operator is a thermal state with an inverse temperature $\beta$, the full Green's functions read (for details, see, e.g., the textbook~\cite{Bellac:2011kqa})\footnote{\label{footnote:M_app}
Here we implicitly assumed that Green's functions of the environment oscillators $Y_n$ are approximated by the free theory propagator at finite temperature. In other words, we assumed that the dynamics of the system oscillator $X$ does not affect properties (e.g., temperature) of the environment thermal bath.
}
\begin{align}
\label{Green_11}
G_{11}(\omega;\omega_n^2)&={\bf P}\,\frac{i}{\omega^2-\omega_n^2}
+2\pi\,{\rm sgn}(\omega)\delta(\omega^2-\omega_n^2)\left[\frac{1}{2}+n_{\rm B}(\omega)\right]\,,
\\
\label{Green_12}
G_{12}(\omega;\omega_n^2)&=2\pi\,{\rm sgn}(\omega)\delta(\omega^2-\omega_n^2)n_{\rm B}(\omega)\,,
\\
\label{Green_21}
G_{21}(\omega;\omega_n^2)&=2\pi\,{\rm sgn}(\omega)\delta(\omega^2-\omega_n^2)\left[1+n_{\rm B}(\omega)\right]\,,
\\
\label{Green_22}
G_{22}(\omega;\omega_n^2)&=(G_{11}(\omega;\omega_n^2))^*\,,
\end{align}
where ${\bf P}$ denotes the principal value, ${\rm sgn}(\omega)$ is the sign function, and $n_{\rm B}(\omega)$ is the Bose-Einstein distribution $n_{\rm B}(\omega)=(e^{\beta\omega}-1)^{-1}$. Notice that the non-zero off-diagonal components only contribute on-shell $\omega=\omega_n$. This is essentially why we need the SK effective action, rather than the in-out effective action, for effective descriptions of dissipations and noises associated with on-shell particle creation. Also, for the same reason, the effective theory after coarse graining is in general non-unitary with a complex effective action. 

\subsection{Effective theory}
\label{subsec:EFT}

\paragraph{Generality.}

Now let us discuss properties of the effective theory~\eqref{EFT_XY}:
\begin{align}
\label{EFT_osci}
iS_{\rm eff}=-\frac{1}{2}\sum_{s,s'=1,2}\int\frac{d\omega}{2\pi}
\left[
X_s(-\omega)D_{ss'}(\omega;\omega_0^2)X_{s'}(\omega)
+X_s(-\omega)\cG_{ss'}(\omega)X_{s'}(\omega)
\right]\,,
\end{align}
where we defined the effective coupling matrix $\cG_{ss'}(\omega)$ by
\begin{align}
\cG_{ss'}(\omega)=\sum_{n=1}^Ng_n^2G_{ss'}(\omega;\omega_n^2)\,.
\end{align}
In general, $\cG_{ss'}(\omega)$ is non-analytic with respect to $\omega$ because of the delta functions, i.e., the on-shell contributions, in the Green's functions~\eqref{Green_11}-\eqref{Green_22}. Note that Eqs.~\eqref{Green_11}-\eqref{Green_22} are for the initial thermal state, but it is a general statement that non-analyticity appears through on-shell contributions. This non-analyticity creates a non-local effect in the effective action~\eqref{EFT_osci}. This non-locality is especially prominent if the number of the environment oscillators, $N$, is not very large. In the field theory intuition, this simply means that the effect of particle productions in the environment sector appears as non-local effects in the effective description of the system.

\paragraph{Local effective theory.}

On the other hand, this non-local effect can be smoothed out when $N$ is large, for which it is convenient to introduce a spectral function $J(\omega_*^2)$ such that 
\begin{align}
\cG_{ss'}(\omega)=\int_0^\infty d\omega_*^2J(\omega_*^2)G_{ss'}(\omega;\omega_*^2)
\quad
{\rm with}
\quad
J(\omega_*^2)=\sum_{n=1}^Ng_n^2\delta (\omega_*^2-\omega_n^2)\,.
\end{align}
Note that the sum over $n$ can be approximated by an integral if $N$ is large, hence non-analyticity may be resolved depending on the distributions of $(\omega_n^2,g_n)$. While the spectral function $J(\omega_*)$ depends on details of the environment sector and its coupling to the system sector, the effective theory indeed becomes local if the spectral function ${\rm sign}(\omega_*)J(\omega_*^2)$ is analytic at $\omega_*=0$. For the so-called Ohmic bath modeling such situations, the effective coupling matrix $\cG_{ss'}(\omega)$ is approximated for small $\omega$ as follows\footnote{We performed renormalization of the frequency $\omega_0$ of the system oscillator $X$ to simplify presentation.}  (see, e.g., \cite{Kamenev,Hongo:2018ant} for details):
\begin{align}
\cG_{11}(\omega)\simeq 2\gamma T\,,
\quad
\cG_{12}(\omega)\simeq -\gamma\omega+2\gamma T\,,
\quad
\cG_{21}(\omega)\simeq \gamma\omega+2\gamma T\,,
\quad
\cG_{22}(\omega)\simeq 2\gamma T\,,
\end{align}
where we have assumed that the initial density operator is a thermal state with a temperature $T=1/\beta$. Also, positivity of $J(\omega_*^2)$ requires $\gamma>0$. The corresponding effective action in the position space is
\begin{align}
\nonumber
iS_{\rm eff}[X_1,X_2]&=\frac{i}{2}\int dt\bigg[
\left(\dot{X}_1^2-\omega_0^2X_1^2\right)-\left(\dot{X}_2^2-\omega_0^2X_2^2\right)
\\*
\label{EFT_osci_12}
&\qquad\qquad\quad
-\gamma(X_1\dot{X_2}-\dot{X}_1X_2)+2i\gamma T(X_1^2+X_2^2-2X_1X_2)
\bigg]\,.
\end{align}
As advertised, we find local effective interactions that mix $X_1$ and $X_2$ after integrating out the environment oscillators, which originate from the on-shell part of the Green's functions.

\paragraph{$r$-$a$ basis.}

For a better physical interpretation of the effective action~\eqref{EFT_osci_12}, it is convenient to define a new pair of variables $X_r$ and $X_a$ by  
\begin{align}
\label{r-a_basis}
X_r=\frac{X_1+X_2}{2}\,,
\quad
X_a=X_1-X_2\,.
\end{align}
Physically speaking, $X_r$ describes the physical profile of $X$, whereas $X_a$ captures fluctuations. In the following, we call $(X_1,X_2)$ the $1$-$2$ basis and $(X_r,X_a)$ the $r$-$a$ basis. In the $r$-$a$ basis, the effective action~\eqref{EFT_osci_12} reads
\begin{align}
\label{EFT_osci_ra}
iS_{\rm eff}[X_r,X_a]&=i\int dt
\left[
-(\ddot{X}_r+\gamma\dot{X}_r+\omega_0^2X_r)X_a+i\gamma TX_a^2
\right]\,.
\end{align}
The equations of motion (EOMs) for $X_r$ and $X_a$ imply
\begin{align}
\ddot{X}_r+\gamma\dot{X}_r+\omega_0^2X_r=0\,,
\quad
X_a=0\,,
\end{align}
which reproduce the Brownian motion with a dissipation coefficient $\gamma$. Also, in the computation of correlation functions, the last term in Eq.~\eqref{EFT_osci_ra} gives Gaussian random noise. In this manner, dissipation and noise are induced by integrating out the environment sector.

\subsection{Consistency of the effective theory}
\label{subsec:consistency}

It is worth noticing that the effective action~\eqref{EFT_osci_12} enjoys the following properties:
\begin{align}
\label{consistency_1}
&S_{\rm eff}[X_1=X,X_2=X]=0\,,
\\
&S_{\rm eff}[X_2,X_1]=-\left(S_{\rm eff}[X_1,X_2]\right)^*\,,
\\
&{\rm Im}\,S_{\rm eff}[X_1,X_2]\geq 0\,,
\end{align}
or equivalently, in the $r$-$a$ basis,
\begin{align}\label{consistency_4}
&S_{\rm eff}[X_r,X_a=0]=0\,,
\\
&S_{\rm eff}[X_r,-X_a]=-\left(S_{\rm eff}[X_r,X_a]\right)^*\,,
\\
\label{consistency_6}
&{\rm Im}\,S_{\rm eff}[X_r,X_a]\geq 0\,.
\end{align}
These are actually generic properties of the SKEFT that follow from unitarity of the microscopic theory and the self-adjointness of the initial density operator. To see this, it is convenient to go back to the definition~\eqref{W_op} of the closed-time-path generating functional $W(J_1,J_2)$ as we will elaborate below.

\medskip
First, unitarity of the microscopic theory implies that the time-evolution operator $U_J$ in Eq.~\eqref{W_op} satisfies
\begin{align}
U_J^\dagger(t_2,t_1) U_J(t_2,t_1)=1\,,
\end{align}
which leads to the following condition on the closed-time-path generating functional:
\begin{align}
W[J_1=J,J_2=J]=0\,.
\end{align}
Also, norm positivity together with the Cauchy-Schwarz inequality applied to Eq.~\eqref{W_op} leads to
\begin{align}
{\rm Im}\,W[J_1,J_2]\geq0\,,
\end{align}
which simply means that probability decreases after tracing out positive-norm states. Finally, the self-adjointness $\rho_0^\dagger=\rho_0$ implies
\begin{align}
W[J_1,J_2]=-(W[J_2,J_1])^*\,.
\end{align}
The conditions~\eqref{consistency_1}-\eqref{consistency_6} follow in the same manner, essentially because we can think of the system variables are kept fixed like the sources when integrating out the environment variables.

\paragraph{Comments on the Kubo-Martin-Schwinger (KMS) condition.}

In the present analysis, we have assumed that the initial state is thermal, for which the correlation functions are related to each other, e.g., as Eqs.~\eqref{Green_11}-\eqref{Green_22} show. This condition is known as the KMS condition. This relation in turns correlates coefficients of the dissipation term (the second term) and the noise term (the last term) in the effective action~\eqref{EFT_osci_ra}, hence the name fluctuation-dissipation relation (FDR). Although the KMS condition is typically presented as a condition for the generating functional, it can be generalized to a symmetry transformation acting on the dynamical fields at the action level. From this symmetry prospective, the FDR can be captured by requiring the action to be invariant under the transformations,
\begin{align}
X_r'(t)=X_r(-t)\,,
\quad
X_a'(t)=X_a(-t)+i\beta\partial_{-t}X_r(-t)\,.
\end{align}
This symmetry of the effective action is called the dynamical KMS symmetry~\cite{Glorioso:2017fpd,Liu:2018kfw,Sieberer:2015hba}. More details on its properties will be explained in Appendix~\ref{HydroEFT}. We emphasize that this symmetry is a consequence of the choice of the initial state, so it is not a universal property of the SKEFT. Indeed, if we choose other initial states, coefficients of the dissipation and noise terms enjoy other types of relations reflecting the properties of the Green's functions in the environment sector.

\subsection{General properties of the Schwinger-Keldysh EFT}
\label{subsec:general}

Various properties of the harmonic oscillator model explained in the previous subsections hold in field theories too. Below we summarize general properties of the Schwinger-Keldysh EFT (see, e.g., a review article~\cite{Liu:2018kfw} for details):
\begin{enumerate}
\item Dynamical variables, collectively denoted by $\chi$, are described by a pair of path-integral variables $(\chi_1,\chi_2)$, reflecting the closed-time-path nature of the formalism.

\item The EFT is in general non-unitary. In particular, the non-unitarity is due to mixing of the $1,2$ sectors, which originates from on-shell particle creations in the environment sector and results in, e.g., dissipation and noise effects. 

\item Consistency such as unitarity of the microscopic theory behind the EFT implies
\begin{align}
\label{consistency_EFT_1}
S_{\rm eff}[\chi_1=\chi,\chi_2=\chi]=0\,,
\,\,
S_{\rm eff}[\chi_2,\chi_1]=-(S_{\rm eff}[\chi_1,\chi_2])^*\,,
\,\,
{\rm Im}\,S_{\rm eff}[\chi_1,\chi_2]\geq 0\,.
\end{align}
Similarly to Eq.~\eqref{r-a_basis}, it is convenient to define
\begin{align}
\chi_r=\frac{\chi_1+\chi_2}{2}\,,
\quad
\chi_a=\chi_1-\chi_2\,.
\end{align}
In this $r$-$a$ basis, the conditions~\eqref{consistency_EFT_1} are rephrased as
\begin{align}
S_{\rm eff}[\chi_r,\chi_a=0]=0\,,
\,\,
S_{\rm eff}[\chi_r,-\chi_a]=-(S_{\rm eff}[\chi_r,\chi_a])^*\,,
\,\,
\label{consistency_EFT_6}
{\rm Im}\,S_{\rm eff}[\chi_r,\chi_a]\geq 0\,.
\end{align}

\item The SKEFT depends not only on the Hamiltonian of the environment sector, but also details of the initial state, since the mixing of the $1$-$2$ sector arises through on-shell particle creation. For example, if the initial state is thermal, the SKEFT should satisfy the dynamical KMS condition, a generalization of the fluctuation-dissipation relation. This type of information can also be taken into account in the EFT construction, once the initial state is specified.
\end{enumerate}

\section{Diffeomorphism symmetries in gravitational Schwinger-Keldysh EFTs \label{sec:diff_sym}}

In this section we elaborate on diffeomorphism symmetries in the SKEFT. We first introduce the two types of diffeomorphism symmetries corresponding to the doubled path-integral variables in the SK formalism in Sec.~\ref{subsec:microscopic}, based on the microscopic description. We then discuss their impacts on the SKEFT in Sec.~\ref{subsec:diffs_mixing}.

\subsection{Microscopic description}
\label{subsec:microscopic}

For illustration, let us consider a scalar field $\phi$ coupled to a dynamical metric $g_{\mu\nu}$ and an environment sector described by microscopic dynamical fields collectively denoted by $\Sigma$: 
\begin{align}
\label{micro_single}
S[\phi,g_{\mu\nu},\Sigma]=\int d^4x\sqrt{-g}
\left[
\frac{M_{\rm Pl}^2}{2}R-\frac{1}{2}(\partial_\mu\phi)^2-V(\phi)+\mathcal{L}_{\rm env}[\phi,g_{\mu\nu},\Sigma]
\right]\,,
\end{align}
where $M_{\rm Pl}$ is the reduced Planck mass and $\mathcal{L}_{\rm env}$ is the microscopic Lagrangian density describing the dynamics of the environment fields $\Sigma$ as well as their couplings to $\phi$ and $g_{\mu\nu}$. In the language of the path-integral form~\eqref{generating_function_PI} of the SK formalism, we have $\Phi=\phi$, $g_{\mu\nu}$, $\Sigma$ as the microscopic dynamical fields. For simplicity, we switch off the source terms as they are not important in the present paper, but it is straightforward to reintroduce them if necessary. We assume that the gravitational microscopic action~\eqref{micro_single} enjoys the diffeomorphism symmetries. We emphasize that this setup is chosen for illustrative purposes, but the following discussion is not restricted to it. The discussion remains universal as long as the underlying microscopic theory is general covariant.

\paragraph{Doubled diffeomorphism symmetries.}

The corresponding SK action, i.e., the exponent of the path-integral weight in Eq.~\eqref{generating_function_PI} is factorized as
\begin{align}
S_{\rm micro}[\Phi_1,\Phi_2]=S[\Phi_1]-S[\Phi_2]\,,
\end{align}
where we have introduced a pair of path-integral variables $\Phi_s=\phi_s$, $g_{s\mu\nu}$, $\Sigma_s$ ($s=1,2$). This implies that $S_{\rm micro}$ enjoys two types of diffeomorphism symmetries, each of which acts on $\Phi_1$ and $\Phi_2$ separately. More concretely, we define the first type of diffeomorphism symmetries that we call diffs$_1$ by the following transformation rule:
\begin{subequations}
\label{eqn:diff12}
\begin{align}
\phi_1'(x)&=\phi_1(x-\xi_1(x))\,, \label{eqn:diffphi1}
\\
g'_{1\mu\nu}(x)&=\frac{\partial (x-\xi_1(x))^\rho}{\partial x^\mu}\frac{\partial (x-\xi_1(x))^\sigma}{\partial x^\nu}g_{1\rho\sigma}(x-\xi_1(x))\,, \label{eqn:diffg1}
\\
\phi_2'(x)&=\phi_2(x)\,, \label{eqn:diffphi2}
\\
g'_{2\mu\nu}(x)&=g_{2\mu\nu}(x) \,, \label{eqn:diffg2}
\end{align}
\end{subequations}
with a local transformation parameter $\xi_1^\mu(x)$, and similarly for the environment fields $\Sigma_s$. Note that the first set of variables $\Phi_1$ transforms in the standard manner, whereas the second set $\Phi_2$ is unchanged. We also define the second type of diffeomorphism symmetries that we call diffs$_2$ in a similar fashion which only acts non-trivially on $\Phi_2$ but $\Phi_1$.

\paragraph{Comments on boundary conditions.}

The microscopic action $S_{\rm micro}$ in the SK formalism enjoys the doubled diffeomorphism symmetries, schematically written as
\begin{align}
\label{diffs_SK}
{\rm diffs}_1\times{\rm diffs}_2\,,
\end{align}
since the action is factorized into the $\Phi_1$ sector and the $\Phi_2$ sector. In contrast, the boundary conditions imposed at the initial time $t_i$ and the final time $t_f$ of the SK path-integral mix the two sectors. For example, the boundary conditions at $t=t_f$ are
\begin{align}
\Phi_1(t_f,\bsx)=\Phi_2(t_f,\bsx)\,,
\end{align}
which break the doubled diffeomorphism symmetries on the boundary into the diagonal one satisfying
\begin{align}
\xi_1^\mu(t_f,\bsx)=\xi_2^\mu(t_f,\bsx)\,.
\end{align}
In a similar fashion, boundary conditions at the initial time break the doubled diffeomorphism symmetries. However, we emphasize that the two diffeomorphism symmetries in the bulk ($t_i<t<t_f$) are not affected by the boundary conditions, thanks to the local nature of diffeomorphism symmetries.

\subsection{Schwinger-Keldysh EFT}
\label{subsec:diffs_mixing}

We have explained that in the SK formalism, the microscopic theory enjoys the doubled diffeomorphism symmetries (except at $t=t_i,t_f$) thanks to the local nature of the symmetries. As a consequence, the SKEFT after integrating out the environment sector should also respect the doubled diffeomorphism symmetries. This is in sharp contrast to global symmetries. On the other hand, as we discussed in Sec.~\ref{sec:SK_review}, the SKEFT accommodates mixing of the doubled path-integral variables $\Phi_1$ and $\Phi_2$, as a consequence of on-shell particle creation that results in dissipation and noise effects. Actually, these mixing effects make the SKEFT construction somewhat non-trivial as we explain below.

\paragraph{Impacts of mixing effects.}

For illustration, let us imagine constructing a SK effective action of a pair of scalars $\phi_s$ and metrics $g_{s\mu\nu}$ ($s=1,2$). The SKEFT accommodates mixing of the $1,2$ sectors, so that one may consider, e.g., the following operator as a candidate for the effective coupling:
\begin{align}
\int d^4x\sqrt{-g_1}\phi_1\phi_2\,.
\end{align}
Obviously, this explicitly breaks the doubled diffeomorphism symmetries ${\rm diffs}_1\times{\rm diffs}_2$ with the transformation parameters $\xi_1^\mu\neq\xi_2^\mu$, while the diagonal part $\xi_1^\mu=\xi_2^\mu$ is still preserved. More generally, we find that local effective couplings mixing the $1,2$ sectors break the non-diagonal diffeomorphism symmetries, as long as we use fields in linear representations of diffeomorphisms only. In other words, mixing couplings require St\"uckelberg fields that non-linearly realize the non-diagonal diffeomorphism symmetries. In Secs.~\ref{sec:diss_sca_grav}-\ref{sec:diss_grav}, we will illustrate this property more explicitly with concrete examples.

\paragraph{Physical and noise diffs.}

Having said this, we reorganize the doubled diffeomorphism symmetries ${\rm diffs}_1\times{\rm diffs}_2$ into diagonal and non-diagonal parts. First, we introduce the physical diffeomorphism symmetries with the transformation parameter $\xi^\mu(x)$ by
\begin{align}
\xi_1^\mu(x)=\xi^\mu(x)
\,,
\quad
\xi_2^\mu(x)=\xi^\mu(x)
\,,
\end{align}
under which the $1,2$ variables transform in the same manner. On the other hand, we introduce the noise diffeomorphism symmetries with the transformation parameter $\xi_a^\mu(x)$ by the non-diagonal part of ${\rm diffs}_1\times {\rm diffs}_2$ such that
\begin{align}
\xi_1^\mu=\frac{1}{2}\xi_a^\mu(x)
\,,
\quad
\xi_2^\mu=-\frac{1}{2}\xi_a^\mu(x)\,,
\end{align}
under which the $1,2$ variables transform in the opposite way.

\paragraph{$r$-$a$ basis.}

To discuss the physical meaning of these symmetries, it is convenient to introduce the $r$-$a$ variables as
\begin{align}
\Phi=\frac{\Phi_1+\Phi_2}{2}
\,,
\quad
\Phi_a=\Phi_1-\Phi_2\,.
\end{align}
Here and in what follows, we suppress the subscript $r$ of $r$-variables for visual clarity. In particular, we define the $r$-$a$ variables for the metrics by
\begin{align}
g_{\mu\nu}=\frac{g_{1\mu\nu}+g_{2\mu\nu}}{2}
\,,
\quad
g_{a\mu\nu}=g_{1\mu\nu}-g_{2\mu\nu} \,,
\label{eqn:gr_ga}
\end{align}
with lower indices and we raise and lower the indices by the physical metric $g_{\mu\nu}$ and its inverse $g^{\mu\nu}$.
Under the physical diffeomorphism transformations, the $r$-$a$ variables for tensor fields transform simply as tensors, e.g.,
\begin{align}
\phi'(x)&=\phi(x-\xi(x))
\,,
\\
\phi'_a(x)&=\phi_a(x-\xi(x))
\,,
\\
g'_{\mu\nu}(x)&=\frac{\partial (x-\xi(x))^\rho}{\partial x^\mu}\frac{\partial (x-\xi(x))^\sigma}{\partial x^\nu}g_{\rho\sigma}(x-\xi(x))
\,,
\\
g'_{a\mu\nu}(x)&=\frac{\partial (x-\xi(x))^\rho}{\partial x^\mu}\frac{\partial (x-\xi(x))^\sigma}{\partial x^\nu}g_{a\rho\sigma}(x-\xi(x))\,.
\label{eqn:phy_diff}
\end{align}
On the other hand, the noise diffeomorphism transformations of the scalars are
\begin{align}
\phi'(x)&=
\frac{\phi(x_+)+\phi(x_-)}{2}
+\frac{\phi_a(x_+)-\phi_a(x_-)}{4}\,, \label{eqn:phirNdiff}
\\
\phi_a'(x)&=
\frac{\phi_a(x_+)+\phi_a(x_-)}{2}
+
\left[
\phi(x_+)-\phi(x_-)
\right]\,, \label{eqn:phiaNdiff}
\end{align}
where $x_\pm^\mu=x^\mu\mp\frac{1}{2}\xi_a^\mu(x)$. Similarly, the transformation rule of the metrics reads
\begin{align}
\nonumber
g_{\mu\nu}'(x)
&
=
\frac{1}{2}
\left[
\frac{\partial x_+^\rho}{\partial x^\mu}\frac{\partial x_+^\sigma}{\partial x^\nu}g_{\rho\sigma}(x_+)
+
\frac{\partial x_-^\rho}{\partial x^\mu}\frac{\partial x_-^\sigma}{\partial x^\nu}g_{\rho\sigma}(x_-)
\right]
\\*
&\quad
+
\frac{1}{4}
\left[
\frac{\partial x_+^\rho}{\partial x^\mu}\frac{\partial x_+^\sigma}{\partial x^\nu}g_{a\rho\sigma}(x_+)
-
\frac{\partial x_-^\rho}{\partial x^\mu}\frac{\partial x_-^\sigma}{\partial x^\nu}g_{a\rho\sigma}(x_-)
\right]
\,, \label{eqn:grNdiff}
\\
\nonumber
g_{a\mu\nu}'(x)
&
=
\frac{1}{2}
\left[
\frac{\partial x_+^\rho}{\partial x^\mu}\frac{\partial x_+^\sigma}{\partial x^\nu}g_{a\rho\sigma}(x_+)
+
\frac{\partial x_-^\rho}{\partial x^\mu}\frac{\partial x_-^\sigma}{\partial x^\nu}g_{a\rho\sigma}(x_-)
\right]
\\*
&\quad
+
\left[
\frac{\partial x_+^\rho}{\partial x^\mu}\frac{\partial x_+^\sigma}{\partial x^\nu}g_{\rho\sigma}(x_+)
-
\frac{\partial x_-^\rho}{\partial x^\mu}\frac{\partial x_-^\sigma}{\partial x^\nu}g_{\rho\sigma}(x_-)
\right]
\,. \label{eqn:gaNdiff}
\end{align}

\subsection{Classical limit}\label{semiclassical}

The noise diffeomorphism transformations \eqref{eqn:phirNdiff}-\eqref{eqn:gaNdiff} in terms of the $r$-$a$ variables look rather complicated, but they are simplified in the classical limit $\hbar\to 0$, where quantum fluctuations are negligible compared to the noise effects originating from the coupling to the environment sector. To see this, let us assign the scale $\hbar$ to the $a$-variables as
\begin{align}
\Phi_r\to\Phi_r\,,\quad \Phi_a \to \hbar \Phi_a\,,
\end{align} 
and define the classical limit by $\hbar\to0$ while keeping $\Phi_r$\ and $\Phi_a$ fixed\footnote{
One might wonder that fluctuations around classical solutions disappear in the classical limit, since $\Phi_1-\Phi_2=\hbar\Phi_a$ goes to zero. However, it is not the case. To see this, let us consider the harmonic oscillator model~\eqref{EFT_osci_ra} for illustration. Fist, recovering $\hbar$ gives the effective action of the form,
\begin{align}
\frac{i}{\hbar}S_{\rm eff}=
\frac{i}{\hbar}\int dt
\left[
-X_a(\ddot{X}_r+\gamma\dot{X}_r+\omega_0^2X_r)+i\gamma \frac{T}{\hbar}X_a^2
\right]\,,
\end{align}
where the temperature $T$ is replaced as $T\to T/\hbar$ as the dimensional analysis implies. Redefining the $a$-variable $X_a$ as $X_a\to \hbar X_a$, we have
\begin{align}
\frac{i}{\hbar}S_{\rm eff}=
i\int dt
\left[
-X_a(\ddot{X}_r+\gamma\dot{X}_r+\omega_0^2X_r)+i\gamma TX_a^2
\right]\,,
\end{align}
so that the fluctuation term, i.e., the second term survives even in the classical limit $\hbar\to0$ ($T$ fixed).
}. Correspondingly, we assign $\hbar$ to the noise diffeomorphism transformation parameter $\xi_a^\mu$ as
$
\xi_a^\mu\to\hbar \xi_a^\mu
$
and take the limit $\hbar\to0$ while keeping $\xi_a^\mu$ fixed. Then, the noise diffeomorphism transformations~\eqref{eqn:phirNdiff}-\eqref{eqn:gaNdiff} are simplified in the classical limit as
\begin{subequations}
\label{eqn:noisediffphig}
\begin{align}
\phi'(x)&=\phi(x)
\,,
\\
\phi_a'(x)&=\phi_a(x)-\xi_a^\mu\partial_\mu\phi(x)\,,\label{noisescalar}
\\
\label{physicaldiffeo}
g_{\mu\nu}'(x)
&=g_{\mu\nu}(x)
\,,
\\
g_{a\mu\nu}'(x)
&=g_{a\mu\nu}(x)-\nabla_\mu\xi_{a\nu}-\nabla_\nu\xi_{a\mu}
\,.\label{noisediffeo}
\end{align}
\end{subequations}
Here we emphasize that the above is for a finite transformation, rather than an infinitesimal one. Also, if needed, it is straightforward to perform the semiclassical expansion to include higher orders in $\hbar$, which corresponds to expansion in the number of $a$-variables.

\section{Dissipative scalar coupled to dynamical gravity \label{sec:diss_sca_grav}}

In this section, we perform the EFT construction for a dissipative scalar coupled to dynamical gravity, as an illustrative example. In Sec.~\ref{subsec:background}, we first explain how the dissipative effects are closely tied to the noise diffeomorphism symmetry. Then, in Sec.~\ref{sec:warmup}, we demonstrate that dynamical St\"{u}ckelberg fields for this symmetry and the energy-momentum tensor of the environment are required to describe dissipation in the presence of dynamical gravity.
As described in Sec \ref{sec:SK_review}, degrees of freedom in the SK formalism always come in a pair. The partner of the noise diffeomorphism St\"{u}ckelberg field is that for the physical diffeomorphism, which naturally appears in HydroEFT as slow-variables. We then use HydroEFT to model the environment in Sec.~\ref{sec:EFT}.
While Secs.~\ref{subsec:background}-\ref{sec:EFT} focus on the SK action at the linear order in $a$-variables that describe classical EOMs, extension to higher orders in $a$-variables capturing noise effects is discussed in Sec.~\ref{subsec:noise}. Finally, in Sec.~\ref{sec:open}, we quantify impacts of fluctuations in the environment sector on the scalar dynamics and demonstrate that they are negligible when the background energy of the environment is large enough compared to the typical energy of fluctuations (hence we call it the decoupling regime).

\subsection{Dissipative scalar coupled to background metric}
\label{subsec:background}

First, we consider a scalar field $\phi$ coupled to a non-dynamical background metric $g_{\mu\nu}$ to illustrate that the noise diffeomorphism symmetry can be utilized to distinguish dissipative operators from non-dissipative operators in the SKEFT.

\paragraph{Non-dissipative scalar.}

As we explained in Sec.~\ref{sec:SK_review}, the SK action for non-dissipative systems is simply a difference of two copies of the in-out action. For a scalar $\phi$ coupled to a background metric $g_{\mu\nu}$, we have
\begin{align}
S_{\rm n.d.}&=\int d^4x\left[
\sqrt{-g_1}\left(-\frac{1}{2}\partial_\mu \phi_1 \partial^\mu\phi_1 -V(\phi_1)\right)
-\sqrt{-g_2}\left(-\frac{1}{2}\partial_\mu \phi_2 \partial^\mu\phi_2 -V(\phi_2)\right)
\right]\,,
\end{align}
which enjoys the two diffeomorphism symmetries given in \eqref{eqn:diff12} (and similarly for the one acting non-trivially on $\phi_2$ and $g_2$). In the $r$-$a$ basis, the corresponding action in the classical limit reads
\begin{align}
\label{dissipative_scalar_ra}
S_{\rm n.d.}&=\int d^4x\sqrt{-g}\left[\frac{1}{2}T^{(\phi)}_{\mu\nu}g_a^{\mu\nu}
+\left(\Box\phi-V'(\phi)\right)\phi_a\right] \,,
\end{align}
where $V'(\phi)$ is the derivative of $V(\phi)$ with respect to $\phi$ and $T_{\mu\nu}$ is the energy-momentum tensor given by
\begin{align}
T^{(\phi)}_{\mu\nu}&=\partial_\mu\phi\partial_\nu\phi
-\left[\frac{1}{2}\partial_\rho\phi\partial^\rho\phi+V(\phi)\right]g_{\mu\nu}\,.
\end{align}
In the following, unless otherwise stated, we take the classical limit for simplicity. Note that the variation in $\phi_a$ reproduces the standard EOM,
\begin{align}
\Box\phi-V'(\phi)=0\,.
\end{align}
Also, the noise diffeomorphism transformation \eqref{eqn:noisediffphig} of the action gives
\begin{align}
\label{DeltaS_nd}
\Delta S_{\rm n.d.}=
\int d^4x\sqrt{-g}\left[\nabla^\mu T^{(\phi)}_{\mu\nu}
-\left(\Box\phi-V'(\phi)\right)\partial_\nu\phi\right]\xi_a^\nu=0\,,
\end{align}
where we performed integration by parts at the second equality. This expression manifests that the noise diffeomorphism invariance of the action guarantees that the energy-momentum tensor $T^{(\phi)}_{\mu\nu}$ is conserved on-shell.

\paragraph{Dissipative scalar.}

Next, we introduce dissipative effects to the action \eqref{dissipative_scalar_ra}. The SK action now reads
\begin{align}
\label{S_phi}
S_{\phi}=
\int d^4x\sqrt{-g}\left[\frac{1}{2}T^{(\phi)}_{\mu\nu}g_a^{\mu\nu}
+\left(\Box\phi-\gamma u^\mu\partial_\mu\phi-V'(\phi)\right)\phi_a\right]\,,
\end{align}
where $u^\mu$ is a time-like unit vector ($u_\mu u^\mu=-1$) that specifies the time-direction for dissipation and $\gamma$ is the dissipation coefficient. The EOM for $\phi_a$ of the modified action is
\begin{align}
\Box\phi-\gamma u^\mu \partial_\mu\phi-V'(\phi)=0\,,
\end{align}
which describes a scalar with dissipation now. The noise diffeomorphism transformation of the modified action is
\begin{align}
\label{Delta_phi_1}
\Delta S_{\phi}
&=\int d^4x\sqrt{-g}\left[\xi_a^\nu\nabla^\mu T^{(\phi)}_{\mu\nu}
-\left(\Box\phi-\gamma u^\mu\partial_\mu\phi-V'(\phi)\right)\xi_a^\mu\partial_\mu\phi\right]
\\*
\label{Delta_phi_2}
&=
\int d^4x\sqrt{-g}\,\xi_a^\nu\gamma u^\mu\partial_\mu\phi\partial_\nu\phi\,,
\end{align}
which shows that the new dissipation term breaks the noise diffeomorphism invariance of the action, as expected from the general argument of the previous section. Note that one can explicitly show that the noise diffeomorphism symmetry cannot be recovered by any modification of the energy-momentum tensor $T_{\mu\nu}^{(\phi)}$. Also, Eqs.~\eqref{Delta_phi_1}-\eqref{Delta_phi_2} show that the energy-momentum tensor $T^{(\phi)}_{\mu\nu}$ is no longer conserved due to violation of the noise diffeomorphism symmetry:
\begin{align}
\label{conservation_violation}
\nabla^\mu T_{\mu\nu}^{(\phi)}=\partial_\nu\phi\left[\Box\phi-V(\phi)\right]\overset{\rm eom}{=}\gamma u^\mu \partial_\mu\phi \partial_\nu\phi\,,
\end{align}
where $\overset{\rm eom}{=}$ means equality that holds on-shell, i.e., after using EOMs. In this manner, the noise diffeomorphism is useful in distinguishing dissipative terms from non-dissipative ones.

\subsection{Lesson from a naive model with dynamical gravity}
\label{sec:warmup}

Next we elaborate on subtleties in coupling dissipative systems with dynamical gravity. For this, let us consider the following naive model where the dissipative scalar of Eq.~\eqref{dissipative_scalar_ra} is simply coupled to the Einstein gravity\footnote{\label{footnote:grav_action}
In the $1$-$2$ basis, the SK action for the pure Einstein gravity is given by
\begin{align}
S_{\rm pure}=\int d^4x\left[
\sqrt{-g_1}\left(\frac{1}{2}M_{\rm Pl}^2R[g_1]-\Lambda\right)
-\sqrt{-g_2}\left(\frac{1}{2}M_{\rm Pl}^2R[g_2]-\Lambda\right)
\right]\,,
\end{align}
where $R[g_{s}]$ ($s=1,2$) is the Ricci scalar constructed from the metric $g_{s \mu\nu}$ and $\Lambda$ is the cosmological constant. In the classical limit, the corresponding action in the $r$-$a$ basis reads
\begin{align}
\label{pure_G}
S_{\rm pure}=\frac{1}{2}\int d^4x\sqrt{-g}
\left(-M_{\rm Pl}^2G_{\mu\nu}-\Lambda g_{\mu\nu}
\right)g^{\mu\nu}_a
\,.
\end{align}
Here and in what follows we suppress metric indication when curvature tensors are constructed from the $r$-variable metric $g_{\mu\nu}$, e.g., $G_{\mu\nu}=G_{\mu\nu}[g]$. Up to the two-derivative order, Eq.~\eqref{pure_G} indeed gives the general SK action of $g_{\mu\nu}$ and $g_{a\mu\nu}$ that is invariant under both the physical and noise diffeomorphism symmetries. Eq.~\eqref{naive_scalar} is obtained by coupling this to  the model~\eqref{dissipative_scalar_ra} ($\Lambda$ is absorbed into the potential $V(\phi)$).
}:
\begin{align}
\label{naive_scalar}
S_{\rm naive}
&=
\int d^4x\sqrt{-g}\left[\frac{1}{2}\left(-M_{\rm Pl}^2G_{\mu\nu}+T^{(\phi)}_{\mu\nu}\right)g_a^{\mu\nu}
+\left(\Box\phi-\gamma u^\mu\partial_\mu\phi-V'(\phi)\right)\phi_a
\right]\,,
\end{align}
where $G_{\mu\nu}$ is the Einstein tensor constructed from the $r$-variable metric $g_{\mu\nu}$ and $M_{\rm Pl}$ is the reduced Planck mass.

\paragraph{Noise diffs and St\"{u}ckelberg trick.}

As we discussed in the previous section, both of the physical and noise diffeomorphism symmetries have to be respected in the presence of dynamical gravity. On the other hand, the dissipation term in Eq.~\eqref{naive_scalar} breaks the noise diffeomorphism symmetry. A simple trick to recover this symmetry is to introduce the St\"{u}ckelberg fields $X_a^\mu$ that nonlinearly realize the noise diffeomorphism symmetry. Practically, referring to \eqref{noisescalar} and \eqref{noisediffeo}, it is achieved by performing the replacement,
\begin{align}\label{Stuckelberg}
g_{a\mu\nu}\to \cG_{a\mu\nu}\equiv g_{a\mu\nu}+\nabla_\mu X_{a\nu}+\nabla_\nu X_{a\mu}\,,
\quad
\phi_a\to \varphi_a\equiv\phi_a+X_a^\mu\partial_\mu\phi\,, \end{align}
and assigning a non-linear transformation rule of the noise diffeomorphism to $X_a^\mu$ as
\begin{align}
X^\mu_a\to X'^\mu_a=X^\mu_a+\xi_a^\mu(x)\,.
\end{align} 
Note that $X_a^\mu$ transforms as a vector under the physical diffeomorphism. By construction, $\cG_{a \mu\nu}$ and $\varphi_a$ are noise diffeomorphism invariant. Then, the SK action after performing the St\"{u}ckelberg trick reads
\begin{align}
\label{naive_scalar_stu}
S_{\rm naive}
&=
\int d^4x\sqrt{-g}\left[\frac{1}{2}\left(-M_{\rm Pl}^2G_{\mu\nu}+T^{(\phi)}_{\mu\nu}\right)\cG_a^{\mu\nu}
+\left(\Box\phi-\gamma u^\mu\partial_\mu\phi-V'(\phi)\right)\varphi_a
\right]\,,
\end{align}
which is invariant under both of the physical and noise diffeomorphisms. Note that the action~\eqref{naive_scalar} can be thought of as an action in the unitary gauge, in which the St\"{u}ckelberg fields $X_a^\mu$ are eaten by the noise metric $g_{a\mu\nu}$.

\paragraph{Necessity of the environment sector.}

Now let us take a closer look at the action~\eqref{naive_scalar_stu} to show that the naive model fails to describe dissipation. For this, it is convenient to rewrite Eq.~\eqref{naive_scalar_stu} into the following form:
\begin{align}
S_{\rm naive}
&=
\int d^4x\sqrt{-g}\,\bigg[\frac{1}{2}\left(-M_{\rm Pl}^2G_{\mu\nu}+T^{(\phi)}_{\mu\nu}\right)g_a^{\mu\nu}
\nonumber
\\
\label{S_pre2}
&\qquad\qquad\qquad\quad
+\left(\Box\phi-\gamma u^\mu\partial_\mu\phi-V'(\phi)\right)\phi_a
-\gamma u^\mu\partial_\mu\phi\partial_\nu\phi X_a^\nu
\bigg]\,,
\end{align}
where we just preformed integration by parts similarly to the second equality of Eq.~\eqref{DeltaS_nd}. This expression manifests that only the $\gamma$ term requires the St\"{u}ckelberg field $X_a^\mu$ to recover the noise diffeomorphism symmetry. The EOMs for $g_{a\mu\nu}$ and $\phi_a$ give the Einstein equation and the dissipative scalar EOM:
\begin{align}
M_{\rm Pl}^2G_{\mu\nu}=T_{\mu\nu}^{(\phi)}
\,,
\quad
\Box\phi-\gamma u^\mu\partial_\mu\phi-V'(\phi)=0\,.
\end{align}
On the other hand, the EOMs for $X_a^\mu$ are
\begin{align}
\label{EOM_X_a}
\gamma u^\mu\partial_\mu\phi\partial_\nu\phi=0\,.
\end{align}
Note that this can be obtained also by taking divergence of the Einstein equation and applying the dissipative scalar EOM, so that the unitary gauge action~\eqref{naive_scalar} leads to the same result. Notably, Eq.~\eqref{EOM_X_a} requires that the dissipated energy quantified in Eq.~\eqref{conservation_violation} has to be zero, hence dissipation cannot be captured by this naive model.

\medskip
The physical origin of this consequence is simple: Gravity couples to all degrees of freedom universally, so that the energy-momentum tensor appearing in the Einstein equation has to take into account the whole matter sector including the environment. However, the current naive model does not contain the environment sector, hence the energy of the scalar cannot escape anywhere. It is also technically important to notice that Eq.~\eqref{EOM_X_a} does not contain second-time-derivative of $\phi$, so that it gives a constraint equation due to which initial conditions cannot be taken arbitrarily. In particular, Eq.~\eqref{EOM_X_a} requires $u^\mu\partial_\mu \phi=0$, so that time-evolution of the scalar $\phi$ is prohibited. In other words, $X_a^\mu$ play a role of auxiliary fields, which is a technical reason why this naive model failed to describe dissipation.

\paragraph{Summary of the lesson.}

To summarize, we need to take into account the energy-momentum tensor of the environment to describe dissipation in the presence of dynamical gravity, simply because gravity couples to all degrees of freedom universally. For the dissipative scalar, this is achieved by improving the action~\eqref{naive_scalar_stu} into the following form:
\begin{align}
\label{general_scalar_gravity}
S
&=
\int d^4x\sqrt{-g}\left[\frac{1}{2}\left(-M_{\rm Pl}^2G_{\mu\nu}+T^{(\phi)}_{\mu\nu}+T^{(\rm env)}_{\mu\nu}\right)\cG_a^{\mu\nu}
+\left(\Box\phi-\gamma u^\mu\partial_\mu\phi-V'(\phi)\right)\varphi_a
+\ldots
\right]\,,
\end{align}
where $r$-variables of the model include those slow variables in the environment sector collectively denoted by $\Phi$, in addition to $\phi$ and $g_{\mu\nu}$. Also, $a$-variables of the model are $\phi_a$, $g_{a\mu\nu}$, $X_a^\mu$ and possibly other slow variables required to describe the environment sector collectively denoted by $\Phi_a$. $T^{(\rm env)}_{\mu\nu}$ is the energy-momentum tensor of the environment sector that is composed of $g_{\mu\nu}$, $\Phi$, and possibly $\phi$ as well. The time-like unit vector $u^\mu$ and the dissipation coefficient $\gamma$ may have $\Phi$-dependence reflecting properties of the environment. The dots stand for terms linear in $\Phi_a$ as well as $\mathcal{O}(\phi_a)$ terms for interactions between the scalar and the environment that are not displayed in the above. In particular, the EOMs for $g_{a\mu\nu}$ and $X_a^\mu$ are
\begin{align}
M_{\rm Pl}^2G_{\mu\nu}=T^{(\phi)}_{\mu\nu}+T^{(\rm env)}_{\mu\nu}
\,,
\quad
\nabla^\mu\left(T^{(\phi)}_{\mu\nu}+T^{(\rm env)}_{\mu\nu}\right)=0\,,
\end{align}
which can capture the energy transfer from the scalar sector to the environment and also gravity sourced by the environment sector appropriately. In the next subsection, we argue that the EFT of hydrodynamics offers a simple and natural model for the environment.

\subsection{Hydro model for the environment}
\label{sec:EFT}

Now we take into account the environment to which energy of the scalar can escape. As we discussed, the main technical issue of the naive model~\eqref{S_pre2} was that the St\"{u}ckelberg fields $X_a^\mu$ for the noise diffeomorphism played a role of auxiliary fields leading to the constraint equation~\eqref{EOM_X_a} that forbids dissipation. A simple and natural way to resolve this issue is to introduce $r$-variables associated with $X_a^\mu$ and a kinetic term to give them dynamics. Based on the motivation explained below, we model the environment sector by HydroEFT to construct a self-consistent SKEFT of a dissipative scalar coupled to dynamical gravity.

\paragraph{Motivation for HydroEFT.}

Recall that the physical and noise diffeomorphisms are $r$-$a$ components of the doubled diffeomorphisms. Then, it is natural to introduce $r$-variables associated with $X_a^\mu$ as St\"{u}ckelberg fields for the physical diffeomorphism. Notably, these St\"{u}ckelberg fields for the doubled diffeomorphism symmetries are used in HydroEFT~\cite{Glorioso:2016gsa,Glorioso:2017fpd} as the slow-variables that represent hydrodynamic modes. We review this formalism in Appendix~\ref{HydroEFT} to make the paper self-contained (see also the original papers~\cite{Glorioso:2016gsa,Glorioso:2017fpd} and the lecture note~\cite{Liu:2018kfw}), but below we give a minium review useful for following the main text.

\paragraph{Minimum review on HydroEFT.}

First, the SK action of HydroEFT takes the following general form at the linear order in $a$-variables:
\begin{align}
S_{\rm hydro}= \frac{1}{2}\int d^4x\sqrt{-g}
\,T_{\mu\nu}^{\rm (hydro)}\cG^{\mu\nu}_a \,,
\end{align}
where $\cG_{a\mu\nu}$ is defined by Eq.~\eqref{Stuckelberg}. The dynamical $a$-variables of the SKEFT are the St\"{u}ckelberg fields $X_a^\mu$ for the noise diffeomorphism and also the noise metric $g_{a\mu\nu}$ is introduced as an external field. $T_{\mu\nu}^{\rm (hydro)}$ represents the energy-momentum tensor of fluids, whose details will be explained shortly. The EOM for $X_a^\mu$ guarantees the conservation law:
\begin{align}
\nabla^\mu T_{\mu\nu}^{\rm (hydro)}=0\,.
\end{align}
The energy-momentum tensor $T_{\mu\nu}^{(\rm hydro)}$ is constructed from dynamical $r$-variables $\sigma^A(x)$\footnote{Note that although the HydroEFT is more naturally formulated in the fluid spacetime with the associated pair of St\"{u}ckelberg fields $\left(X^\mu(\sigma) , X^\mu_a(\sigma)\right)$, it is more convenient for our purpose to rewrite the hydro action in the physical spacetime, $x^\mu$, the conversion is provided by the inversion map of $X^\mu(\sigma)$, i.e. $\sigma^A(x)$. All the other dynamical fields, e.g. $X^\mu_a(\sigma(x))$ and $\beta(\sigma(x))$, are expressed as functions of $\sigma^A(x)$.\label{foot:inv}}
($A=\bar{0},\bar{1},\bar{2},\bar{3}$) that describe a coordinate map from the physical spacetime (Eulerian) coordinates $x^\mu$ to the fluid spacetime (Lagrangian) coordinates $\sigma^A$, as well as $g_{\mu\nu}$. 

\medskip
For example, a concrete form of the energy-momentum tensor for ideal fluids is
\begin{align}
\label{T_ideal_fluid}
T_{\mu\nu}^{\rm (hydro)}
=T_{\mu\nu}^{(0)}
\equiv
\rho_0 u_\mu u_\nu+p_0\Delta_{\mu\nu}
\quad
(\text{ideal fluids})
\,,
\end{align}
where $u^\mu$ is a time-like unit vector representing the fluid velocity defined by
\begin{align}
\label{def_u}
u^\mu=b^{-1}K^\mu_{\bar{0}}
\quad(u^\mu u_\mu=-1)\,.
\end{align}
Here, $K^\mu_A(x)\equiv \left(\partial_\mu \sigma^{A}(x)\right)^{-1}$ is the inverse of $\partial_\mu \sigma^{A}$ (such that $K^\mu_A \partial_\mu \sigma^{B}=\delta_A^B$) with $\partial_\mu \equiv \frac{\partial \ }{\partial x^\mu}$ and $b = \sqrt{-g_{\mu\nu}K^\mu_{\bar{0}} K^\nu_{\bar{0}}}$. The energy density $\rho_0$ and the pressure $p_0$ are functions of $\beta \equiv b \beta_0$ with a reference inverse temperature $\beta_0=T^{-1}_0$. When fluids are locally in thermal equilibrium, $\beta$ is identified as the local inverse temperature, even though the EFT construction is not limited to the locally thermal case. Higher derivative corrections to fluid dynamics can also be incorporated in a systematic manner.

\medskip
As a final remark, we summarize the doubled diffeomorphisms in HydroEFT and explain why $\sigma^A(x)$ is the variable used in HydroEFT expressed in physical spacetime. First, the physical diffeomorphism transformation of the St\"{u}ckelberg field $X^\mu(\sigma)$ (dynamical variable in the fluid spacetime) (See Footnote \ref{foot:inv} and Appendix \ref{HydroEFT}) is given by
\begin{align}
X^\mu(\sigma')=X^\mu(\sigma)-\xi^\mu(X(\sigma))\,,
\end{align}
where $X^\mu(\sigma)$ is the inverse map of $\sigma^A(x)$ (dynamical variable in the physical spacetime). The connection between $\sigma^A(x)$ and $X^\mu(\sigma)$ shows that $\sigma^A(x)$ play a role of St\"{u}ckelberg fields for the physical diffeomorphism in physical spacetime, as advertised.
On the other hand, we require $\sigma^A(x)$ to be invariant under the noise diffeomorphism similarly to other $r$-variables. This is the reason why we say that HydroEFT offers a simple and natural framework to model the environment sector\footnote{
We are not claiming that HydroEFT is required as a general model for the environment sector, even though it offers a simple and natural framework to construct a local effective action in the presence of dissipation. It would be interesting to explore other possibilities too.}. See Appendix~\ref{HydroEFT} for more details of the EFT construction, including the symmetry principle thereof.

\paragraph{Dissipative scalar coupled to gravity and hydro.}

Now we use HydroEFT to model the environment sector of a dissipative scalar coupled to dynamical gravity. It is simply achieved by considering the following SK action:
\begin{align}
S&=
\int d^4x\sqrt{-g}\left[\frac{1}{2}\left(-M_{\rm Pl}^2G_{\mu\nu}+T^{(\phi)}_{\mu\nu}+T_{\mu\nu}^{(\rm hydro)}\right)\cG_a^{\mu\nu}
+\left(\Box\phi-\gamma u^\mu\partial_\mu\phi-V'(\phi)\right)\varphi_a
\right]\,,
\nonumber
\\
&=
\int d^4x\sqrt{-g}\,\bigg[\frac{1}{2}\left(-M_{\rm Pl}^2G_{\mu\nu}+T^{(\phi)}_{\mu\nu}\right)g_a^{\mu\nu}
+\frac{1}{2}T_{\mu\nu}^{(\rm hydro)}\cG_a^{\mu\nu}
\nonumber
\\
&\qquad\qquad\qquad \ \ \ \,
+\left(\Box\phi-\gamma u^\mu\partial_\mu\phi-V'(\phi)\right)\phi_a
-\gamma u^\mu\partial_\mu\phi\partial_\nu\phi X_a^\nu
\bigg]\,,
\end{align}
where dynamical variables are $\phi$, $\phi_a$, $g_{\mu\nu}$, $g_{a\mu\nu}$, $\sigma^A$, and $X_a$. $u^\mu$ in the dissipation term of $\phi$ is identified with the one defined in Eq.~\eqref{def_u}. Also, it is compatible with the symmetry of HydroEFT to promote the dissipation coefficient~$\gamma$ to a function of $\beta$, just as $\rho_0$ and $p_0$ in Eq.~\eqref{T_ideal_fluid}\footnote{It is also compatible with the symmetry to include an operator $\nabla_\mu u^\mu\varphi_a$ that is at the same order in derivatives as the dissipation term, but we ignore it for simplicity.}.
The EOMs for $g_a^{\mu\nu}$ and $\phi_a$ are
\begin{align}
M_{\rm Pl}^2G_{\mu\nu}=T_{\mu\nu}^{(\phi)}+T_{\mu\nu}^{({\rm hydro})}
\,,
\quad
\Box\phi-\gamma u^\mu\partial_\mu\phi-V'(\phi)=0\,.
\end{align}
On the other hand, the EOM for $X_a^\mu$ is modified as
\begin{align}
\nabla^\mu T_{\mu\nu}^{({\rm hydro})}=-\gamma u^\mu\partial_\mu\phi\partial_\nu\phi\,,
\end{align}
which captures the energy flow from the scalar sector to the environment sector modeled by HydroEFT.
In the next subsection we include noise effects described by higher orders terms in $a$-variables. Also, in Sec.~\ref{sec:open}, we use our model to identify the condition under which fluctuations in the environment sector are irrelevant to the scalar dynamics.

\subsection{Noise effects}
\label{subsec:noise}

So far, we focused on the $\mathcal{O}(a)$ part of the SK action that describes classical EOMs, where the non-trivial features of dynamical gravity already appear at this order. However, it is straightforward to include higher order terms in $a$-variables that describe noise effects in the same fashion. For illustration, we consider coupling a dissipative scalar $\phi$ with the following SK action to dynamical gravity taking into account the environment:
\begin{align}
S_{\phi}=
\int d^4x\sqrt{-g}\left[\frac{1}{2}T^{(\phi)}_{\mu\nu}g_a^{\mu\nu}
+\left(\Box\phi-\gamma u^\mu\partial_\mu\phi-V'(\phi)\right)\phi_a+\frac{i}{2}A\phi_a^2\right]\,.
\end{align}
Here we have added the last term to the earlier action~\eqref{S_phi}, which is second order in $a$-variables and captures Gaussian noise effects in the same manner as the harmonic oscillator model discussed in Sec.~\ref{sec:SK_review}. Note that $A\geq0$ follows from the unitarity condition~\eqref{consistency_EFT_6}.

\paragraph{Generality.}

We begin by a general argument without specifying details of the environment. For a general environment, the SK action \eqref{general_scalar_gravity} at the first order in $a$-variables can be extended to the second order as
\begin{align}
S
&=
\int d^4x\sqrt{-g}\left[\frac{1}{2}\left(-M_{\rm Pl}^2G_{\mu\nu}+T^{(\phi)}_{\mu\nu}+T^{(\rm env)}_{\mu\nu}\right)\cG_a^{\mu\nu}
+\left(\Box\phi-\gamma u^\mu\partial_\mu\phi-V'(\phi)\right)\varphi_a
+\ldots\right.
\nonumber
\\
\label{general_phi_a2}
&
\qquad\qquad\qquad\,\,
\left.
+\frac{i}{2}A\varphi_a^2+\frac{i}{4}W_{\mu\alpha,\nu\beta}\cG_a^{\mu\nu}\cG_a^{\alpha\beta}+\ldots
\right]\,,
\end{align}
where the second line is the newly added $\mathcal{O}(a^2)$ terms. Here we have added a Gaussian noise term constructed from $\cG_{a\mu\nu}$ parameterized by the coefficient $W_{\mu\alpha,\nu\beta}$, which respects the physical and noise diffeomorphism symmetries and is also at the same order in derivatives as the noise term of the scalar. As we will discuss in the next section, this describes noise effects of gravitational waves. Besides, the dots in the second line stand for $\mathcal{O}(a^2)$ terms that include $a$-variables other than $\phi_a,g_{a\mu\nu},X_a^\mu$, as well as non-Gaussian noise terms coming from higher orders.

\paragraph{Hydro model.}

Now we model the environment by HydroEFT. For illustration, we take into account the first-order corrections to ideal fluids, which are at the same order in derivatives as the noise and dissipation effects in the scalar sector. More explicitly, the SK action for first-order fluids in the Landau frame takes the form~\cite{Glorioso:2016gsa,Glorioso:2017fpd},
\begin{align}
\label{gravity_action}
S_{\rm hydro}= \int d^4x\sqrt{-g}\left[
\frac{1}{2}T_{\mu\nu}^{\rm (hydro)}\cG^{\mu\nu}_a +\frac{i}{4}W _{\mu\alpha,\nu\beta}\cG^{\mu\nu}_a \cG^{\alpha\beta}_a\right]\,,
\end{align}
where the energy-momentum tensor $T_{\mu\nu}^{\rm (hydro)}$ is corrected from the ideal fluid one~\eqref{T_ideal_fluid} as
\begin{align}
\label{T_1st_fluid}
T_{\mu\nu}^{\rm (hydro)}
=T^{(0)}_{\mu\nu}+T^{(1)}_{\mu\nu}
\quad
{\rm with}
\quad
T^{(1)}_{\mu\nu}\equiv
-
2\eta \left(K_{\mu\nu}-\frac{1}{3}g_{\mu\nu}K\right)
-\zeta\Delta_{\mu\nu}K
\,.
\end{align}
Here $K_{\mu\nu}$ is the extrinsic curvature constructed from the fluid velocity vector $u^\mu$ and $K$ is its trace part:
\begin{align}\label{extrinsic}
K_{\mu\nu}=\Delta_{\mu}{}^\rho\nabla_\rho u_\nu\,,
\quad
K=\Delta^{\mu\nu}K_{\mu\nu}\,.
\end{align}
The shear viscosity $\eta$ and the bulk viscosity $\zeta$ are functions of $\beta$. On the other hand, the second term representing noise effects has the same form as the one in the general argument~\eqref{general_phi_a2}. To elaborate on it, let us recall that noise terms can be correlated to dissipation terms, once the environment state is specified. For example, when the fluids are locally in thermal equilibrium, the dynamical KMS symmetry implies that $W_{\mu\alpha, \nu\beta}$ is of the form (see Appendix~\ref{HydroEFT})\footnote{
The round (square) brackets imply that the indices are symmetrized (anti-symmetrized), e.g., as $A_{(\mu\nu)}=\frac{1}{2}(A_{\mu\nu}+A_{\nu\mu})\ (A_{[\mu\nu]}=\frac{1}{2}(A_{\mu\nu}-A_{\nu\mu}))$.
},
\begin{align}
\label{W_def}
W_{\mu\alpha,\nu\beta}=
\frac{2\eta}{\beta}\left(
\Delta_{\alpha(\mu}\Delta_{\nu)\beta}-\frac{1}{3}\Delta_{\mu\nu}\Delta_{\alpha\beta}
\right)+\frac{\zeta}{\beta}\Delta_{\mu\nu}\Delta_{\alpha\beta}\,,
\end{align}
with $\beta$ identified with the local inverse temperature. Note that the dynamical KMS symmetry gives a first-law type constraint $\rho_0+p_0=-\beta\frac{\partial  p_0}{\partial \beta}$ on the ideal fluid part as well.

\medskip
Combining it with the general discussion, we obtain the following SK action for a dissipative scalar coupled to dynamical gravity and the environment modeled by HydroEFT: 
\begin{align}
S
&=
\int d^4x\sqrt{-g}\left[\frac{1}{2}\left(-M_{\rm Pl}^2G_{\mu\nu}+T^{(\phi)}_{\mu\nu}+T^{(\rm hydro)}_{\mu\nu}\right)\cG_a^{\mu\nu}
+\left(\Box\phi-\gamma u^\mu\partial_\mu\phi-V'(\phi)\right)\varphi_a
\right.
\nonumber
\\
&
\qquad\qquad\qquad\,\,
\left.
+i\frac{\gamma}{\beta}\varphi_a^2+\frac{i}{4}W_{\mu\alpha,\nu\beta}\cG_a^{\mu\nu}\cG_a^{\alpha\beta}
\right]\,,
\end{align}
where $T^{(\rm hydro)}_{\mu\nu}$ and $W_{\mu\alpha,\nu\beta}$ are defined in Eq.~\eqref{T_1st_fluid} and Eq.~\eqref{W_def}, and $\gamma$ is now a function of the local inverse temperature $\beta$. For simplicity, we have assumed that the environment fluids are locally in thermal equilibrium by imposing the dynamical KMS symmetry, which fixes the noise term of the scalar as $A=2\gamma/\beta$. Relaxing this assumption is straightforward, but there appear some more new terms in the EFT that are not presented here.

\subsection{Decoupling regime \label{sec:open}}

As we discussed, it is necessary to take into account the energy-momentum tensor of the environment sector appropriately, in order to describe dissipative systems in the presence of dynamical gravity. An advantage of the EFT constructed in this manner is that one may study impacts of fluctuations in the environment sector on the system sector of interest, even though it requires modeling of the environment sector. For example, the hydro model of the environment can be used to quantify how the dynamics of the system sector (the scalar $\phi$ in the present model) is affected, e.g., by temperature fluctuations.

\medskip
On the other hand, there are certain situations where fluctuations in the environment sector are irrelevant to the dynamics of the system sector. For example, in the harmonic oscillator model of Sec.~\ref{sec:SK_review}, Green's functions of the environment oscillators $Y_n$ were approximated by the free theory propagator at a fixed finite temperature (see footnote \ref{footnote:M_app}) and the resultant effective theory~\eqref{EFT_osci_ra} was nothing but the one for Brownian motion used in the context of open systems as the simplest example. Intuitively, this Markovian type approximation is justified for Brownian motion because energy of the Brownian particle is too small to change properties (e.g., temperature) of the environment. At the cost of limitation of applicability, this type of approximation has an advantage in universality, i.e., the effective theory of the system sector does not depend on details of the environment. In the following, we call the regime where such approximation works the decoupling regime.

\paragraph{Decoupling regime in hydro model.}

Having said this, let us use the hydro model for the environment to identify the decoupling regime, where fluctuations in the environment sector are irrelevant. For illustration, let us consider the $\mathcal{O}(a)$ action for a dissipative scalar $\phi$ coupled to gravity and the environment sector described by an ideal fluid:
\begin{align}
S&=
\int d^4x\sqrt{-g}\,\bigg[\frac{1}{2}\left(-M_{\rm Pl}^2G_{\mu\nu}+T^{(\phi)}_{\mu\nu}+\rho_0 u_\mu u_\nu+p_0\Delta_{\mu\nu}\right)\cG_a^{\mu\nu}
\nonumber
\\
\label{p+g+if}
&\qquad\qquad\qquad\quad
+\left(\Box\phi-\gamma u^\mu\partial_\mu\phi-V'(\phi)\right)\varphi_a
\bigg]\,.
\end{align}
Also let us denote the background configurations of $\beta$ and $u^\mu$ by $\bar{\beta}$ and $\bar{u}^\mu$, and their fluctuations by $\delta \beta$ and $\delta u^\mu$, respectively. To identify the decoupling regime, let us estimate impacts of $\delta \beta$, $\delta u^\mu$, and $X_a^\mu$ on the $\phi$ dynamics. For example, the dissipation term can be expanded around the background as\footnote{We assumed that the scalar $\phi$ vanishes at the background level.}
\begin{align}
-\big(\gamma(\bar{\beta})+\gamma'(\bar{\beta})\delta \beta\big)\big(\bar{u}^\mu +\delta u^\mu\big)\,\partial_\mu\phi \,\big(\phi_a+X_a^\nu\partial_\nu\phi\big)\,,
\end{align}
so that $\delta \beta$, $\delta u^\mu$, and $X_a^\mu$ are negligible as long as the following conditions are satisfied:
\begin{align}
\label{open_app}
|\gamma'(\bar{\beta})\delta\beta|\ll \left|\gamma(\bar{\beta})\right|\,,
\quad
|\delta u|\ll 1
\,,
\quad
|X_a \partial\phi|\ll|\phi_a|\,.
\end{align}
Here the first condition says that temperature fluctuations are small enough to not change the dissipation coefficient. The second condition says that the time-direction of dissipation is fixed approximately. The last condition says that noise in the fluid configuration is small enough compared to the typical modulation scale of the scalar. Similar conditions can be derived in the same manner for other terms in Eq.~\eqref{p+g+if} and also for higher order corrections in derivatives and $a$-fields.

\medskip
Finally, let us write down the conditions~\eqref{open_app} in terms of the path-integral variables. As we explained in Sec.~\ref{sec:EFT}, path-integral variables of HydroEFT are $\sigma^A(x)$ and $X_a^\mu$. If we write fluctuations of $\sigma$ as $\pi$, the kinetic term of $(\pi,X_a)$ is schematically of the form,
\begin{align}
(\rho_0+p_0)\,\partial \pi\partial X_a\,,
\end{align}
where for simplicity we assumed that $\beta$-dependence of $\rho_0,p_0$ is small enough. Then, canonically normalized fields $\pi_c$ and $X_{ac}$ can be introduced schematically as
\begin{align}
\pi_c\sim \sqrt{\rho_0+p_0}\,\pi\,,\quad X_{ac}\sim  \sqrt{\rho_0+p_0}\,X_a\,.
\end{align}
Now the conditions~\eqref{open_app} determining the decoupling regime are rephrased as
\begin{align}
|\partial X_{ac}|,|\partial\pi_c|\ll \sqrt{\rho_0+p_0}\,,
\end{align}
where we used $\delta u\sim\partial \pi$ and $\delta \beta\sim \bar{\beta}\,\partial \pi$, and also assumed for simplicity that the $\beta$-dependence of $\gamma$ is not so strong $|\frac{\ln\gamma}{\ln\beta}|\lesssim 1$. By dimensional analysis, we conclude that the decoupling regime is given in terms of the typical energy scale $E$ of fluctuations as
\begin{align}
E^4\ll \rho_0+p_0\,,
\end{align}
which means that energy of fluctuations is small enough compared to the background energy of the environment. This is nothing but the condition intuitively explained earlier using Brownian motion. In this manner, the decoupling regime of environment fluctuations can be quantified by explicitly modeling the environment, which is an advantage of the current formulation.

\section{Dissipative gravity \label{sec:diss_grav}}

In the previous section, we constructed the SKEFT for a dissipative scalar coupled to dynamical gravity. There, it was crucial to introduce the energy-momentum tensor of the environment sector, essentially because gravity couples to all degrees of freedom universally. In particular, we used HydroEFT to model the environment. In this section, we apply the same formalism to dissipative gravity, simply by truncating the scalar field. As an application, we discuss dissipative gravitational waves in Sec.~\ref{subsec:dissipativeGW} and then the generalized second law of black hole thermodynamics in Sec.~\ref{second-law}.

\subsection{Dissipative gravitational wave}
\label{subsec:dissipativeGW}

First we construct a SKEFT for dissipative gravitational waves whose linear EOM around a flat spacetime is schematically of the form, 
\begin{align}
\ddot{h}_{ij}^{\rm TT}+\gamma \dot{h}^{\rm TT}_{ij}-\partial_k^2 h_{ij}^{\rm TT}=0\,. \label{eqn:dis_gravwave}
\end{align}
Here $h_{ij}^{\rm TT}$ is the transverse-traceless (TT) component of metric fluctuations with $i$, $j$, $k$ as the spatial indices and $\gamma$ is the dissipation coefficient. We begin by a naive attempt analogous to the naive model of a dissipative scalar studied in Sec.~\ref{sec:warmup} and then incorporate the environment sector to which the energy of gravitational waves can escape.

\subsubsection{A naive model}

To specify the time-direction of dissipation, let us introduce a time-like unit vector $u^\mu$ ($u_\mu u^\mu=-1$). Using this, we also define a projector $\Delta_{\mu\nu}$ onto the three-dimensional spatial surface orthogonal to $u^\mu$ and the associated extrinsic curvature $K_{\mu\nu}$ by
\begin{align}
\Delta_{\mu\nu}=g_{\mu\nu}+u_\mu u_\nu\,,
\quad
K_{\mu\nu}=\Delta_{\mu}{}^\rho\nabla_\rho u_\nu\,.
\end{align}
At the linear order in perturbations around a flat spacetime with $u^\mu=\delta^\mu_0$, the TT components $h_{ij}^{\rm TT}$ of metric fluctuations are embedded into the Einstein tensor $G_{\mu\nu}$ and the extrinsic curvature $K_{\mu\nu}$ as
\begin{align}
G_{ij}\ni\frac{1}{2}\left(
\ddot{h}_{ij}^{\rm TT}-\partial_k^2 h_{ij}^{\rm TT}
\right)\,,
\quad
K_{ij}\ni\frac{1}{2}\dot{h}_{ij}^{\rm TT}\,.
\end{align}
Therefore, a naive candidate for non-linear EOMs of dissipative gravitational waves is
\begin{align}
\label{dis_gw_eom}
G_{\mu\nu}+\gamma K_{\mu\nu}=0\,,
\end{align}
whose TT components reproduce Eq.~\eqref{eqn:dis_gravwave} at the linear level. The corresponding SK action is
\begin{align}\label{dissi_grav}
S_{\rm pre}= -\frac{M_{\rm Pl}^2}{2}\int d^4x
\sqrt{-g} \left(G_{\mu\nu}+\gamma K_{\mu\nu}
\right)g^{\mu\nu}_a\,.
\end{align}
The $\gamma$ term breaks the noise diffeomorphism symmetry, but it can be recovered by introducing the St\"{u}ckelberg fields $X^\mu_a$ as
\begin{align}\notag
S_{\rm pre}
&= -\frac{M_{\rm Pl}^2}{2}\int d^4x
\sqrt{-g} \left(G_{\mu\nu}+\gamma K_{\mu\nu}\right)\cG_{a}^{\mu\nu}
\\\label{dissi_grav_st}
&= -\frac{M_{\rm Pl}^2}{2}\int d^4x
\sqrt{-g} \left(G_{\mu\nu}g^{\mu\nu}_a+\gamma K_{\mu\nu}\cG_{a}^{\mu\nu}\right)\,,
\end{align}
where $\cG_{a\mu\nu}$ is defined in Eq.~\eqref{Stuckelberg}.

\paragraph{Necessity of the environment sector.}

Similarly to the naive model~\eqref{naive_scalar} of a dissipative scalar, the current model~\eqref{dissi_grav_st} cannot describe dissipative gravitational waves due to absence of the environment energy-momentum tensor. To see this, consider the EOM for $X_a^\mu$, or equivalently the divergence of the EOM~\eqref{dis_gw_eom}:
\begin{align}
\label{div_K}
\nabla_\mu K^{\mu\nu}=0\,.
\end{align}
If we assume that the time-like unit vector $u^\mu$ has no acceleration $u^\nu\nabla_\nu u^\mu=0$,\footnote{This is a covariant generalization of the assumption that $u^\mu$ is a constant vector.} the EOM~\eqref{div_K} is reduced to
\begin{align}
\label{div_K_cons}
\Delta^{\mu\rho}\nabla_\rho K_{\mu\nu}=0\,.
\end{align}
Essentially because the derivative acting on $K_{\mu\nu}$ is projected onto the spatial directions, the left hand side does not contain second-time-derivatives of propagating modes, i.e., $\ddot{h}_{ij}^{\rm TT}$, and time-derivatives of auxiliary fields (lapse and shift), hence Eq.~\eqref{div_K_cons} gives a constraint equation due to which the initial conditions cannot be chosen arbitrarily. Indeed, by explicitly calculating the energy-momentum tensor of linear gravitational waves, one can show that this constraint equation prohibits energy loss of gravitational waves and the current naive model~\eqref{dissi_grav_st} cannot describe dissipative gravitational waves, similarly to the naive model of Sec.~\ref{sec:warmup}.

\subsubsection{Introduction of the environment sector}

\paragraph{Generality.}

Now we take into account the environment sector to which the energy of gravitational waves can escape. For generality, consider a SKEFT that contains $g_{\mu\nu}$, $g_{a\mu\nu}$ and $X_a^\mu$ as well as other $r$- and $a$-variables collectively denoted by $\Phi$ and $\Phi_a$. Also suppose that the SK action up to the first order in $a$-variables is schematically of the form,
\begin{align}
S&=
\frac{1}{2}\int d^4x\sqrt{-g}\left[-M_{\rm Pl}^2
(G_{\mu\nu}+\gamma K_{\mu\nu})
+T_{\mu\nu}^{(\rm env)}\right]\cG_a^{\mu\nu}
+\ldots\,,
\end{align}
where $T_{\mu\nu}^{(\rm env)}$ is the energy-momentum tensor of the environment that is composed of $g_{\mu\nu}$ and $\Phi$. The time-like vector $u^\mu$ defining the extrinsic curvature $K_{\mu\nu}$ and the dissipation coefficient $\gamma$ may have $\Phi$-dependence. The dots stand for terms linear in $\Phi_a$, if any. Then, the EOM for $X_a^\mu$ reads
\begin{align}
\label{eom_env_general}
 \nabla^\mu (\gamma K_{\mu\nu})=\nabla^\mu T^{\rm (env)}_{\mu\nu}\,,
\end{align}
which is no more a constraint equation as long as the right hand side $\nabla^\mu T^{\rm (env)}_{\mu\nu}$ contains second-time-derivatives of $\Phi$ or time-derivatives of auxiliary fields. Physically, this equates the energy loss of gravitational waves (the left hand side) and the energy increase in the environment sector (the right hand side). One may check this statement explicitly by expanding Eq.~\eqref{eom_env_general} around a given background.

\paragraph{Hydro model.}

HydroEFT again offers a simple and natural setup that realizes the above scenario. More concretely, consider the following SK action up to the first order in $a$-variables:
\begin{align}
S&=
\frac{1}{2}\int d^4x\sqrt{-g}\left(-M_{\rm Pl}^2G_{\mu\nu}+T_{\mu\nu}^{(\rm hydro)}\right)\cG_a^{\mu\nu}\,.
\end{align}
Here $T_{\mu\nu}^{(\rm hydro)}$ is the energy-momentum tensor of fluids that is composed of $\sigma^A$ as well as $g_{\mu\nu}$. Its explicit form up to the first order in the Landau frame is
\begin{align}
\label{T_mn_1st_hydro_GW}
T_{\mu\nu}^{(\rm hydro)}=\rho_0 u_\mu u_\nu+p_0\Delta_{\mu\nu}
-
\left[2\eta \left(K_{\mu\nu}-\frac{1}{3}g_{\mu\nu}K\right)+\zeta\Delta_{\mu\nu}K
\right]
\,,
\end{align}
where $\rho_0$, $p_0$, $\eta$, and $\zeta$ are functions of the local inverse temperature $\beta$. Notably, the shear viscosity $\eta$ term contains the extrinsic curvature, which sources dissipation of gravitational waves. For example, if the background fluid is homogeneous and the corresponding energy-momentum tensor is small enough, the background spacetime is approximated as a flat spacetime and $u^\mu=\delta^\mu_0$ in the rest frame. Then, the linear EOM for the TT component of metric fluctuations reproduces Eq.~\eqref{eqn:dis_gravwave} with the dissipation coefficient $\gamma=2\eta/M_{\rm Pl}^2$. Note that fluctuations of the hydro variables $X^\mu$ do not appear in the linear EOM~\eqref{eqn:dis_gravwave} simply because there are no spin-$2$ modes in the hydro sector. However, they appear from the second order of fluctuations, which are responsible for describing the energy flow from the gravitational wave sector to the environment hydro sector and contribute to the non-linear dynamics.

\paragraph{Noise effects.}

To close this subsection, we discuss the noise effects of dissipative gravitational waves, which appear through the effective operators at the second order in $a$-variables. Up to the second order in $a$-variables, the SK action of first-order fluids coupled to dynamical gravity takes the form,
\begin{align}
S&=
\int d^4x\sqrt{-g}
\left[
\frac{1}{2}\left(-M_{\rm Pl}^2G_{\mu\nu}+T_{\mu\nu}^{(\rm hydro)}\right)\cG_a^{\mu\nu}
+\frac{i}{2}W_{\mu\alpha,\nu\beta}\cG_a^{\mu\nu}\cG_a^{\alpha\beta}
\right]
\,.
\label{eqn:gravhydroaction}
\end{align}
In the Landau frame, $T_{\mu\nu}^{(\rm hydro)}$ is given by \eqref{T_mn_1st_hydro_GW}, whereas $W_{\mu\alpha,\nu\beta}$ is
\begin{align}
W_{\mu\alpha,\nu\beta}
=
\frac{2\eta}{\beta}\left(
\Delta_{\alpha(\mu}\Delta_{\nu)\beta}-\frac{1}{3}\Delta_{\mu\nu}\Delta_{\alpha\beta}
\right)+\frac{\zeta}{\beta}\Delta_{\mu\nu}\Delta_{\alpha\beta}\,.
\end{align}
Note that the dynamical KMS condition specific to thermal setups was used to relate the shear viscosity $\eta$ and the bulk viscosity $\zeta$ with the coefficients in $W_{\mu\alpha,\nu\beta}$, but it can be relaxed if necessary. While the $W_{\mu\alpha,\nu\beta}$ term describes noise effects of fluids, it also captures the noise effects of gravitational waves as well. For example, if the background fluid is homogeneous and the spacetime curvature is negligible, the second order SK action for the TT component of metric fluctuations in the fluid rest frame reads
\begin{align}
S=\frac{M_{\rm Pl}^2}{4}\int d^4x
\left[
-\left(\ddot{h}_{ij}^{\rm TT}+\gamma\dot{h}_{ij}^{\rm TT}-\partial_k^2h_{ij}^{\rm TT}\right)h_{aij}^{\rm TT}
+i\frac{\gamma}{\beta}h_{aij}^{\rm TT}h_{aij}^{\rm TT}
\right]
\end{align}
with the dissipation coefficient $\gamma=2\eta/M_{\rm Pl}^2$. We confirm that the dynamical KMS symmetry implies the fluctuation-dissipation relation of dissipative gravitational waves.

\subsection{Generalized second law of black hole thermodynamics}\label{second-law}

In this subsection, we discuss the generalized second law~\cite{Bekenstein:1972tm,Bekenstein:1973ur,Bekenstein:1974ax} in the context of HydroEFT coupled to dynamical gravity introduced in the previous subsection. For simplicity, we focus on first-order fluids coupled to the Einstein gravity~\eqref{eqn:gravhydroaction}, which is free from higher derivative terms with Riemann tensors and follows the area law of black hole entropy.
We consider a scenario where a dynamical black hole is surrounded by a fluid and evaluate the change of generalized entropy within the setup~\eqref{eqn:gravhydroaction}.

\paragraph{Preliminaries.}

For stationary black holes, the event horizon, apparent horizon, and Killing horizon all coincide, hence there is no ambiguity in the area law.
While it is still an ongoing issue on how to extend the notion of the area law to dynamical black holes, a plausible extension is to use the area of the apparent horizon~\cite{Hayward:1993wb,Hayward:1997jp,Ashtekar:2002ag,Nielsen:2005af,Nielsen:2008cr,Debnath:2012fqp,Hollands:2024vbe,Visser:2024pwz,Takeda:2024qbq,Kong:2024sqc}, which can be defined without following the full time-evolution of the black hole, in contrast to the event horizon.
In this context, the temperature of the black hole is defined using the dynamical surface gravity determined by the Kodama vector \cite{Hayward:1997jp,Nielsen:2007ac,Hayward:2008jq}. Meanwhile, the fluid surrounding the black hole has a locally defined temperature, leading to a situation where two distinct temperatures are present. In the following, we show that the generalized entropy is generated by the temperature gradient between the fluid and the black hole, as in standard thermodynamics.

\subsubsection{Dynamical black hole thermodynamics}

First, we review the thermodynamics of dynamical black holes based on Refs.~\cite{Hayward:1997jp,Nielsen:2007ac}.

\paragraph{Metric ansatz.}

We start by considering a spherically symmetric spacetime, which takes the following general form parameterized by two functions $F(v,r)$ and $h(v,r)$:
\begin{align}
d s^2=-e^{2h(v,r)}F (v,r)dv^2+2e^{h(v,r)}dvdr+ r^2 \left(d\theta^2+\sin^2 \theta d\phi^2\right)\,,
\label{eqn:metric}
\end{align} 
where $\theta$ and $\phi$ are the angular coordinates, $r$ is the radial coordinate, and $v$ is the ingoing null coordinate. To describe a dynamical black hole in an asymptotically flat spacetime, we further parameterize $F(v,r)$~as
\begin{align}
F(v,r)=1-\frac{2GM(v,r)}{r}\,,
\quad
M(v,r)>0\,,
\end{align}
and assume that $\displaystyle\lim_{r\to\infty}F(v,r)=1$. Here we introduced the Newton constant $G=\frac{1}{8\pi M_{\rm Pl}^2}$. We also assume that $h(v,r)$ is finite in the spacetime region of our interest. Note that the Vaidya spacetime is its special case with $M(v,r)=M(v)$ and $ h(v,r)=0$, but our setup is not limited to this.

\paragraph{Apparent horizon.}

The location of the apparent horizon is determined by solving $F(v,r)=0$. We introduce its radius $r_{\rm H}(v)$ as a function of $v$ such that
\begin{align}
r_{\rm H}(v)=2GM_{\rm BH}(v)
\quad
{\rm with}
\quad
M_{\rm BH}(v)=M\left(v,r_{\rm H}(v)\right)\,,
\label{eqn:hor}
\end{align}
where $M_{\rm BH}(v)$ is the Misner-Sharp mass~\cite{PhysRev.136.B571}
inside the horizon that we will call it the black hole mass in the following. Note that normalization of the energy is set by the Kodama vector~\cite{Kodama:1979vn}\footnote{
The Kodama vector is defined by 
$
k^\mu\equiv \epsilon^{\mu\nu}\nabla_\nu r$,
where $\epsilon^{\mu\nu}$ is the Levi-Civita tensor on the $2$-dimensional $v$-$r$ spacetime, which is normalized as $\epsilon^{vr}=-\epsilon^{rv}=1/ \sqrt{-\det{q_{\mu\nu}}}$, with $q_{\mu\nu}$ being the induced metric on the 2 dimensional spacetime~\cite{Kodama:1979vn}. }
\begin{align}
\label{killing}
k^\mu(v,r)\partial_\mu=e^{-h(v,r)}\partial_v\,.
\end{align}
With the above definition of the black hole mass, its time-dependence 
reads
\begin{align}
\dot{M}_{\rm BH}(v)&=\dot{M}(v,r_{\rm H}(v))+ M'(v,r_{\rm H}(v))\dot{r}_{\rm H}(v)\,.
\label{eqn:bhmasschange}
\end{align}
Here and in what follows, we use dot and prime to denote derivatives in $v$ and $r$, respectively.
To see the physical meaning of each term, it is convenient to introduce the black hole volume $V_{\rm BH}(v)$ as
\begin{align}
V_{\rm BH}(v)\equiv \frac{4\pi r_{\rm H}(v)^3}{3}\,,
\end{align}
and one can obtain from Eq.~\eqref{eqn:bhmasschange} that
\begin{align}
\dot{M}_{\rm BH}(v)&=\dot{M}(v,r_{\rm H}(v))+ \frac{M'(v,r_{\rm H}(v))}{4\pi r_{\rm H}(v)^2}\dot{V}_{\rm BH}(v)\,.
\label{eqn:BHmass}
\end{align}
We interpret the terms $\dot{M}_{\rm BH}(v),\ \dot{M}(v,r_{\rm H}(v))$ and $\frac{M'(v,r_{\rm H}(v))}{4\pi r_{\rm H}(v)^2}\dot{V}_{\rm BH}(v)$ as the changes of internal energy, heat, and work, respectively. Using Eq.~(\ref{eqn:hor}), the time-dependence of the horizon radius reads
\begin{align}
\dot{r}_{\rm H}(v)=2G\dot{M}_{\rm BH}(v)=\frac{2G\dot{M}(v,r_{\rm H}(v))}{1-2GM'(v,r_{\rm H}(v))}\,.
\end{align}
We also introduce the entropy $S_{\rm BH}(v)$ of the time-evolving black hole proportional to the area of the apparent horizon as
\begin{align}
S_{\rm BH}(v)=\frac{4\pi r_{\rm H}(v)^2}{4G}\,.
\end{align}
Then, its time-dependence reads
\begin{align}
\dot{S}_{\rm BH}(v)
&=\frac{4\pi r_{\rm H}(v)}{1-2GM'(v,r_{\rm H}(v))}
\dot{M}(v,r_{\rm H}(v))
\,. \label{BH-1st-pre}
\end{align}

\paragraph{Black hole temperature.}

The relation~\eqref{BH-1st-pre} motivates us to introduce the inverse temperature $\beta_{\rm BH}(v)$ of the time-evolving black hole by
\begin{align}
\beta_{\rm BH}(v)=\frac{4\pi r_{\rm H}(v)}{1-2GM'(v,r_{\rm H}(v))}\,,
\end{align}
in terms of which Eq.~\eqref{BH-1st-pre} enjoys the following first-law type relation:
\begin{align}
\dot{S}_{\rm BH}(v)=\beta_{\rm BH}(v)\dot{M}(v,r_{\rm H}(v))\,.
\end{align}
Indeed, it is related to the so-called dynamical surface gravity $\kappa(v)$  as\footnote{
We followed the definition~\eqref{dsg_def} proposed in~\cite{Hayward:1997jp} and further studied in~\cite{Nielsen:2007ac,Hayward:2008jq,Murk:2020wkm,Mann:2021lif,Murk:2021cla,Kurpicz:2021kgf,Murk:2023vdw}.
See however  Refs.~\cite{Visser:1992qh,Hayward:2008jq,Abreu:2010ru,Pielahn:2011ra,Cropp:2013zxi} for other approaches to surface gravity in dynamical spacetimes.
}
\begin{align}
\beta_{\rm BH}(v)=\frac{2\pi}{\kappa(v)}=\frac{4\pi r_{\rm H}(v)}{1-2GM'(v,r_{\rm H}(v))}\,,
\end{align}
where the general definition of $\kappa(v)$ is given in terms of the Kodama vector $k^\mu$ by
\begin{align}
\label{dsg_def}
\left.k^\mu\nabla_{[\mu} k_{\nu]}\right|_{r=r_H(v)}=\left.\kappa(v) k_\nu\right|_{r=r_H(v)}\,,
\end{align}
and its concrete form in our setup~\eqref{eqn:metric} reads
\begin{align}
\kappa(v)
=\frac{1-2GM'(v,r_{\rm H}(v))}{4GM(v,r_{\rm H}(v))}
=\frac{1-2GM'(v,r_{\rm H}(v))}{2r_{\rm H}(v)}
\,.
\end{align}

\subsubsection{Generalized second law}

\paragraph{Setup.}

\begin{figure}[t] 
	\centering 
	\includegraphics[width=5cm]{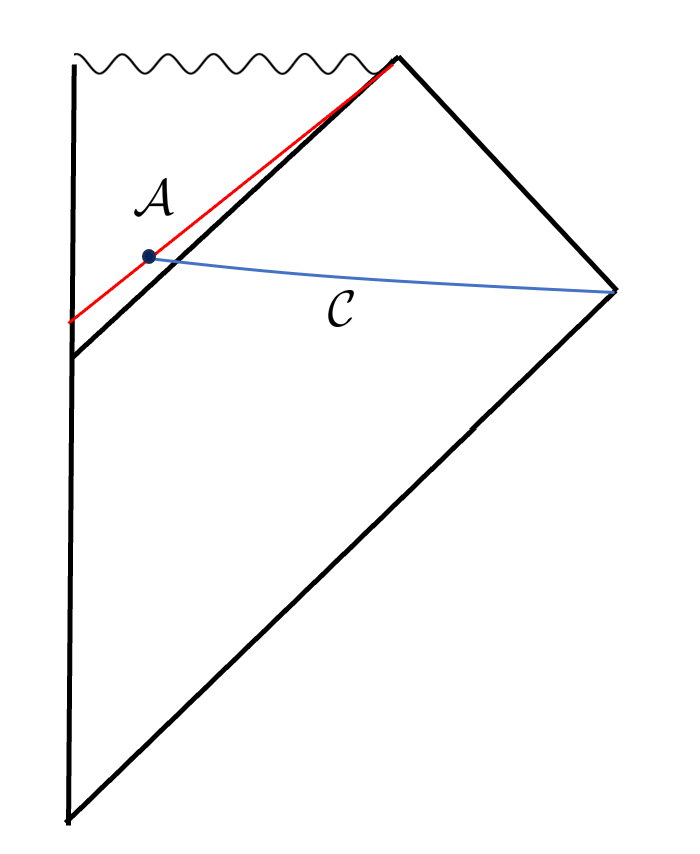} 
	\caption{This figure shows the Penrose diagram of a non-stationary black hole: The black line corresponds to the event horizon and the red line corresponds to the apparent horizon. Here we consider the area of the apparent horizon 
$\cal A$ and the entropy of matter on the time slice $\cal C$.}
\label{fig2}
\end{figure}

Now we suppose that time-evolution of the black hole is induced by a flow of fluid matter and discuss the generalized second law in this context. We parameterize Cauchy surfaces by the coordinate $v$ at its intersection with the apparent horizon. See Fig.~\ref{fig2}.

\paragraph{Energy flow.}

On the apparent horizon, $\dot{M}$ and $M'$ are related to the energy-momentum tensor $T_{\mu\nu}$ of the fluid through the Einstein equation as
\begin{align}
\dot{M}(v,r_{\rm H}(v))=4\pi r_{\rm H}(v)^2e^{-h}T_{vv}
\,,
\quad
M'(v,r_{\rm H}(v))=-4\pi r_{\rm H}(v)^2e^{-h}T_{vr}\,.
\end{align}
Here and in what follows we suppress the superscript ``hydro'' of $T_{\mu\nu}^{(\rm hydro)}$ for visual clarity. Also we refer to $h(v,r_{\rm H}(v))$ and $T_{\mu\nu}(v,r_{\rm H}(v))$ as $h$ and $T_{\mu\nu}$, as long as there is no risk of confusion. Then, time-dependence of the black hole mass $M_{\rm BH}$ from (\ref{eqn:BHmass}) reads
\begin{align}
\nonumber
\dot{M}_{\rm BH}(v)
&=4\pi r_{\rm H}(v)^2\,T_v{}^\mu
\partial_\mu(r-r_{\rm H}(v))\,,
\end{align}
which is nothing but the rate of the total energy flowing across the apparent horizon. 

\paragraph{Fluid entropy.}

We next consider entropy of fluids outside the apparent horizon:
\begin{align}
S_{\rm mat}(\bar{v})
=\int d^4x\sqrt{-g}\,\delta(v-\bar{v})\,\theta(r-r_{\rm H}(v))\,s^v\,,
\end{align}
where $s^\mu\equiv s u^\mu = (\rho_0+p_0)\beta u^\mu$ is the entropy current of the fluid (see, e.g., Refs. \cite{Crossley:2015evo,Glorioso:2016gsa,Glorioso:2017fpd} and Appendix~\ref{HydroEFT}).
Its time-dependence follows from the Stokes' theorem that
\begin{align}
\frac{dS_{\rm mat}(\bar{v})}{d\bar{v}}
&=-\int d^4x\sqrt{-g}\delta'(v-\bar{v})\theta(r-r_{\rm H}(v)) s^v
\nonumber
\\
&=-\int d^4x\sqrt{-g}\,\partial_\mu\delta(v-\bar{v})\,\theta(r-r_{\rm H}(v)) \,s^\mu
\\
\nonumber
&=
\int d^4x\sqrt{-g}\,\delta(v-\bar{v})
\Big[
\theta(r-r_{\rm H}(v)) \,\nabla_\mu s^\mu
+\delta(r-r_{\rm H}(v)) \,s^\mu\partial_\mu(r-r_{\rm H}(v)) 
\Big]\,,
\end{align}
where the local second law $\nabla_\mu s^\mu\geq 0$ (that follows from unitarity and the dynamical KMS symmetry~\cite{Glorioso:2016gsa}) implies that the first term in the square brackets is non-negative. Then, the second term gives a lower bound on the entropy change:
\begin{align}
\frac{dS_{\rm mat}(v)}{dv}
&\geq
4\pi r_{\rm H}(v)^2\,e^{h}
s^\mu\partial_\mu(r-r_{\rm H}(v))\,,
\end{align}
where $h=h(v,r_{\rm H}(v))$ and $s^\mu=s^\mu(v,r_{\rm H}(v))$ as before.
\paragraph{Generalized second law.}

With the entropy change of the black hole and fluid matter introduced in the above, we can now discuss the generalized entropy $S_{\rm gen}$ given by the sum,
\begin{align}
S_{\rm gen}(v)=S_{\rm BH}(v)+S_{\rm mat}(v)\,.
\end{align}
Using the results so far, its time-dependence follows as
\begin{align}\notag
\frac{dS_{\text{gen}}(v)}{dv}& \geq \beta_{\rm BH}(v)\left(\dot{M}_{\rm BH}(v)- \frac{M'(v,r_{\rm H}(v))}{4\pi r_{\rm H}(v)^2}\dot{V}_{\rm BH}(v)\right) + 4 \pi r_{\rm H}(v)^2 e^{h} s^\mu \partial_\mu (r - r_{\rm H}(v))\\
&=4\pi r_{\rm H}^2e^h\left(\beta_{\rm BH}-\beta_{\text{eff}}\right)T_{\mu\nu}k^\mu k^\nu +4\pi r_{\rm H}^2e^h \beta_{\rm eff}\left(T^{(1)}_{\mu\nu}k^\mu k^\nu-T_{\rm eff}s^\mu \partial_\mu r_{\rm H}(v) \right)\,,
\label{gen_ent}
\end{align}
where we have defined the effective temperature of the fluid as
\begin{align}
\beta_{\text{eff}}=\frac{1}{T_{\rm eff}} \equiv-\frac{1}{ (k^\mu u_\mu)T}\,.
\end{align}
Note that the effective temperature does not in general coincide with the local temperature $T$ but differs by a factor $-(k^\mu u_\mu)$, where $k^\mu$ is defined in \eqref{killing}. This correction factor plays a similar role as the blue-shift factor in the definition of Tolman temperature \cite{Hayward:2008jq}. The blue-shift factor and hence the effective temperature vanishes when $k^\mu$ and $u^\mu$ are parallel $k^\mu u_\mu=0$, whereas it becomes larger as the fluid is boosted to the inward direction\footnote{
It is worth noticing that for the Schwarzschild spacetime ($h(v,r)=0$ and $M(v,r)=M$ is a constant), $k^\mu u_\mu=-1$ when taking $u^\mu$ as a velocity of a free-fall particle emitted from infinity, indicating that there is no blue/red-shift. This is consistent with the equivalence principle.}.

\paragraph{Quasi-static approximation.}

While it is not easy to provide a full interpretation of the general result~\eqref{gen_ent}, a certain simplification occurs if we focus on quasi-static processes and neglect terms proportional to $\dot{r}_H(v)$:
\begin{align}
\frac{dS_{\text{gen}}(v)}{dv}& \geq 
4\pi r_{\rm H}^2e^h\left(\beta_{\rm BH}-\beta_{\text{eff}}\right)T_{\mu\nu}k^\mu k^\nu +4\pi r_{\rm H}^2e^h \beta_{\rm eff}T^{(1)}_{\mu\nu}k^\mu k^\nu
\quad
(\text{quasi-static})\,,
\end{align}
where the second term is from the first order corrections to ideal fluids, hence it is subdominant at least when $\beta_{\rm BH}\neq\beta_{\rm eff}$. Interestingly, the sign of the dominant first term is controlled by the temperature gradient between the black hole and fluids, and also the null energy. Below, we discuss its property for the $T_{kk}>0$ case and the $T_{kk}<0$ case separately.

\paragraph{Case 1: $T_{vv}>0$.}
When the matter flow satisfies the null energy condition, the generalized entropy increases when $\beta_{\rm BH}>\beta_{\text{eff}}$. This is consistent with the general thermodynamic expectation that the total entropy of a closed system increases when heat flows from a warmer region to a colder region.

\paragraph{Case 2: $T_{vv}<0$.}

Conversely, when the null energy condition is violated, the generalized entropy increases when $\beta_{\rm BH}<\beta_{\text{eff}}$. A typical example for this is particle emission through the Hawking radiation \cite{Hawking:1975vcx}. Again, it is consistent with thermodynamic expectation that heat flow is controlled by the temperature gradient. Furthermore, for Hawking particles in quasi-static processes, we expect that $u^\mu$ is proportional to $k^\mu$, in which limit the inverse effective temperature goes as $\beta_{\rm eff} \to\infty$ and $\beta_{\rm BH}<\beta_{\rm eff}$ is trivially satisfied.

\section{Summary and discussion \label{sec:summary}}

In this paper we elaborated on the effective field theory (EFT) construction for dissipative open systems coupled to dynamical gravity based on the Schwinger-Keldysh (SK) formalism.
Dynamical gravity theories contain two sets of diffeomorphism symmetries in the SK formalism, reflecting the closed-time-path nature of the formalism. The two diffeomorphism symmetries can be decomposed into the diagonal physical diffeomorphism and the off-diagonal noise diffeomorphism, both of which have to be preserved in the EFT construction. In particular, dissipation is tied to breaking of the noise diffeomorphism symmetry. To recover the symmetry, we performed the St\"{u}ckelberg trick, which enables us to study dissipative effects while preserving the noise diffeomorphism symmetry manifestly.

\medskip
The results above stem from the property that gravity couples to all degrees of freedom. Specifically, energy conservation is realized through the coupling of all degrees of freedom via the St\"{u}ckelberg field associated with the noise diffeomorphism symmetry. It turns out that including only the noise diffeomorphism St\"{u}ckelberg field is not adequate to account for energy loss due to dissipation, and an environmental sector serving as a sink is further required. Although not unique, one such natural and simple choice of an environment can be achieved by introducing the partner St\"{u}ckelberg field (associated with the physical diffeomorphism) of the two diffeomorphism symmetries. The dynamics of the partner St\"{u}ckelberg field describes a fluid which acts as an environment of which energy can escape to and dissipation can occur in other sectors.
Although the total system including all the sectors is a closed system, where energy is conserved but exchange can occurs between different sectors, the system of interest can be regarded as open when the energy of the background energy of the environment is sufficiently large and fluctuations in the environment sector are irrelevant to the system sector dynamics.

\medskip
Although our illustrative open system EFT contained a dissipative scalar matter coupled to the gravity and environment sectors, we can also consider a scenario where the gravity sector is dissipative itself. In Sec. \ref{sec:diss_grav}, we constructed a SKEFT model of dissipative gravitation wave. We identified the origin of the dissipative effect of gravitation wave comes from an extrinsic curvature term. When expanded around the flat spacetime, the SKEFT action reproduces the correct gravitational wave equation including the dissipative effect. As illustrated in Sec. \ref{sec:diss_sca_grav}, an environment sector is essential for the energy of the system of interest to dissipate to (or more accurately, transfer to). It was no different in this case where we introduced the environment sector modeled by a fluid described by HydroEFT.
As a further application of this framework to gravitational theories, this paper utilizes results from the SKEFT of hydrodynamics to discuss a generalized second law in scenarios where a black hole evolves dynamically while surrounded by a fluid. In dynamical black holes, black hole entropy is typically characterized by the apparent horizon. Here, it is shown that generalized entropy is produced due to a temperature gradient arising from the local fluid temperature, introduced based on local equilibrium associated with the dynamical KMS symmetry, and the black hole temperature, determined by spacetime symmetries. Moreover, this temperature gradient is influenced by the blue-shift effect depending on the fluid's direction. The analysis confirms that both (i) in the usual case where fluid flows into the black hole, and (ii) in the case where exotic matter, such as the Hawking radiation, flows out of the black hole, the direction of matter flow and the temperature gradient enhance entropy production.

\medskip
There are various future directions along the line of the present paper. First, it is interesting to apply the dissipative scalar model of Sec.~\ref{sec:diss_sca_grav} to cosmology. As we discussed there, interactions between the scalar sector and the environment sector are present in general. Impacts of these new interactions on dissipative inflation will be worth studying. Second, our EFT recipe can directly be applied to the open EFT of inflation~\cite{Hongo:2018ant,Salcedo:2024smn}. Even though the current studies in the literature are limited to fixed background spacetime, our present work offers a framework to incorporate dynamical gravity and go beyond the decoupling limit of gravity, by which dynamics of gravitational waves in the open EFT of inflation can be studied for example. 
Another important issue is the model-dependence of the environment sector.
Although it is natural and simple to use HydroEFT as a fluid model for the dynamical St\"{u}ckelberg field associated with the physical diffeomorphism symmetry, we have implicitly imposed a set of internal symmetries which defines a fluid. However, the HydroEFT formalism is not limited to this and one can modify the set of symmetries imposed on the St\"{u}ckelberg fields to describe other phases of matter such as solid and superfluid~\cite{Baggioli:2020haa,Landry:2020ire}. It is interesting to utilize the flexibility of the HydroEFT formalism to study various environment sector scenarios and explore a possibility to identify the phase of the environment matter from the dynamics of dissipative open systems. Besides, more fundamentally, the SK formalism is a framework useful to study real-time dynamics at finite temperature, so it will offer a proper arena to study thermodynamic properties of dynamical gravity and spacetime. We hope that our present work on the SKEFT with dynamical gravity gives useful insights for such studies as well.

\acknowledgments
K.N. and T.N. thank Maria Mylova, Perseas Christodoulidis, Sota Sato and  Vaios Ziogas for fruitful discussion during the collaboration on a related topic.
We also thank Keisuke Izumi, Hideo Furugori, Daisuke Yoshida and Kaho Yoshimura for useful discussion.
T.N. is supported in part by JSPS KAKENHI Grant No. JP22H01220 and MEXT KAKENHI Grant No. JP21H05184 and No. JP23H04007. K.N. is supported by JSPS KAKENHI Grant Number JP22J20380. P.H.C.L acknowledges the support from JSPS KAKENHI Grant No.~20H01902 and JP23H01174 and MEXT KAKENHI Grant No. 21H05462.

\appendix

\section{Schwinger-Keldysh EFT of hydrodynamics}\label{HydroEFT}

This appendix reviews and explains all the essential features and ideas of HydroEFT used in this paper. We refer the readers to \cite{Crossley:2015evo,Glorioso:2016gsa,Glorioso:2017fpd,Liu:2018kfw} for proofs and derivations.

\subsection{St\"{u}ckelberg fields of HydroEFT}
\label{subsec:StuHydroEFT}

\begin{figure}[t] 
	\centering 
	\includegraphics[width=9.5cm]{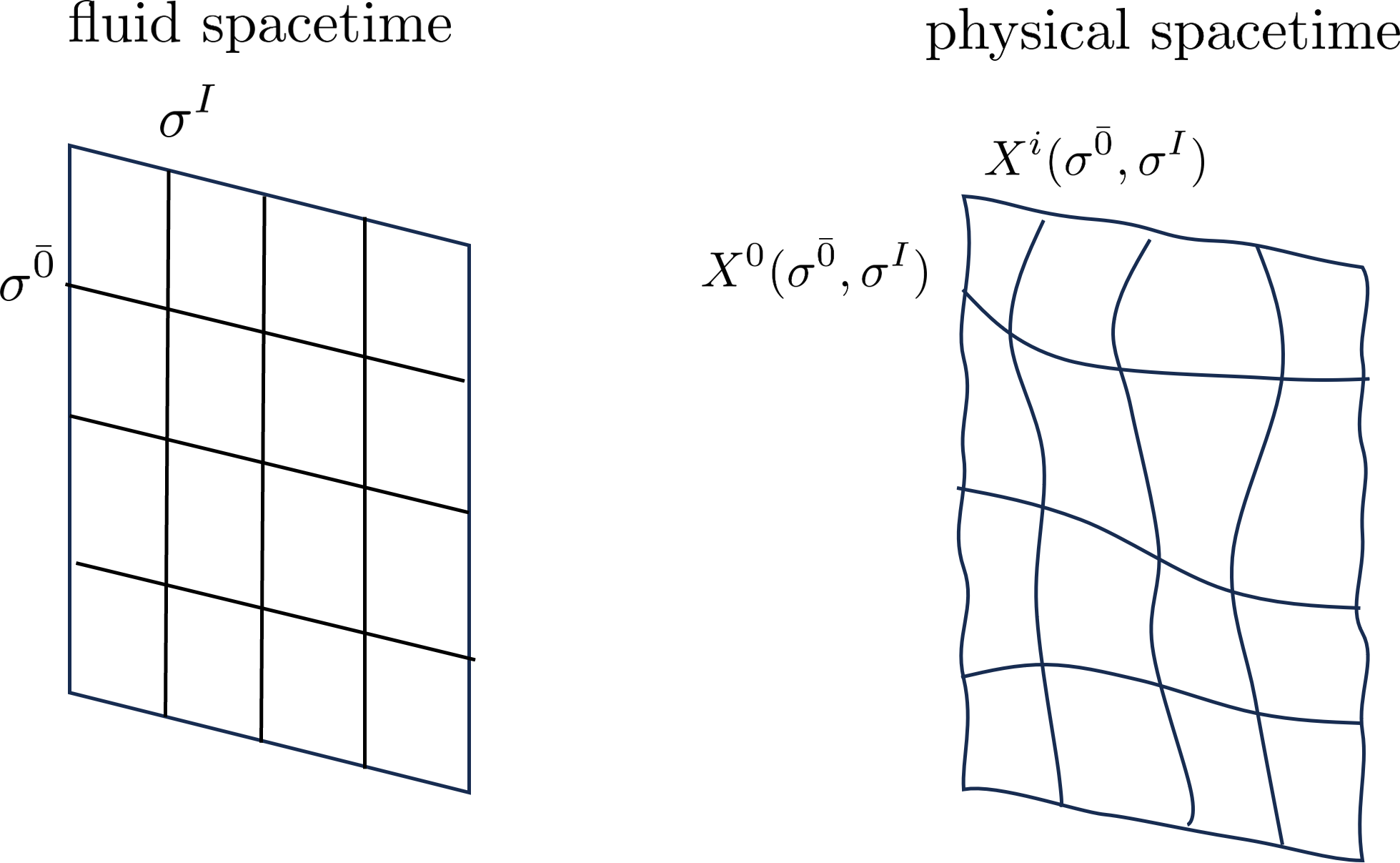}
	\caption{The left figure shows the fluid spacetime where ${\sf h}_{AB}$ lives: Fluid elements are labeled by $\sigma^{\bar{0}}$ and $\sigma^I$ ($I=\bar{1},\bar{2},\bar{3}$). The right figure shows the physical spacetime where $g_{\mu\nu}$ lives: $X^\mu(\sigma)$ tells us the position of the fluids in the spacetime. Note that there are two copies of physical spacetime but only one is shown here. }
\label{fig4}
\end{figure}

In this subsection, we explain the physical origin of the St\"{u}ckelberg fields that arise in Hydro EFT. Recall that in Eq. \eqref{gravity_action}, we presented a hydro action
\begin{align}
S_{\rm hydro}&= \int d^4x\sqrt{-g}\left[
\frac{1}{2}T_{\mu\nu}^{\rm (hydro)}\cG^{\mu\nu}_a +\frac{i}{4}W _{\mu\alpha,\nu\beta}\cG^{\mu\nu}_a \cG^{\alpha\beta}_a\right]\,,
\label{eqn:app_hyaction}
\end{align}
which implicitly contains the dynamical fields $X^\mu$ and $X^\mu_a$ in the $r$-$a$ basis or equivalently $X_1^\mu$ and $X_2^\mu$ in the 1-2 basis. These fields were introduced in Eq.~\eqref{Stuckelberg} through the St\"{u}ckelberg trick to recover the diffeomorphism symmetries at the action level on the closed-time path.
To make a connection to conventional hydrodynamics and discuss the physical meaning of these fields, it is best done by going back to the original form of the action constructed in \cite{Crossley:2015evo,Glorioso:2016gsa}. The exact form of the action is not important here and is schematically written as
\begin{align}
    S_{\text{hydro}}&=\int d^4\sigma\mathcal{L}_{\rm eff}[X_1^\mu(\sigma),X_2^\mu(\sigma);g_{1 \mu\nu},g_{2 \mu\nu}]\,.
\label{eqn:fluspace_act}
\end{align}
The effective Lagrangian, $\mathcal{L}_{\rm eff}$, is written in terms of coordinates $\sigma^A$. In this form, this hydro action is expressed in a ``fluid spacetime'' parameterized by the coordinates $\sigma^A\, (A=\bar{0},\bar{1},\bar{2},\bar{3})$. The dynamical fields $X_1^\mu(\sigma)$/$X_2^\mu(\sigma)$ describes a map from the fluid spacetime to a physical spacetime parameterized by the coordinates $x_1^\mu$/$x_2^\mu$  (see Figure~\ref{fig4}). This is equivalent to the Lagrangian frame of conventional hydrodynamics. Note that we have two copies of the physical spacetime. $g_{1\mu\nu}$ and $g_{2\mu\nu}$ in Eq. \eqref{eqn:fluspace_act} are the two physical spacetime metrics and act as the (non-dynamical) source of the stress tensor living on each physical spacetime, respectively. In this language, the induced metric ${\sf h}_{sAB}$ on the fluid spacetime in the 1-2 basis reads
\begin{align}
{\sf h}_{sAB}(\sigma)= g_{s\mu\nu}(X_s)\frac{\partial X_s^\mu}{\partial \sigma^A}\frac{\partial X_s^\nu}{\partial \sigma^B},\quad (s=1,2)\,.
\label{pullback}
\end{align} 
To convert the hydro action to a more familiar form, i.e., the effective Lagrangian and all the fields are functions of the physical spacetime, formally we can introduce an inverse map of the dynamical variables and express the fluid spacetime coordinates in terms of the physical spacetime coordinates. Such inverse map was presented and discussed in \cite{Glorioso:2017fpd}. 

\subsection{Symmetries and local temperature of HydroEFT}

Having discussed the physical meaning of the dynamical fields of HydroEFT, the next step is to understand the symmetries of a fluid. After that, we also explain how we can introduce a local temperature to the system and how the concept of local thermal equilibrium is connected to the KMS symmetry in this subsection.

\paragraph{Fluid symmetries.}

In the Lagrangian frame of hydrodynamics, we label each fluid element and follows their trajectory (c.f. the Eulerian frame where we focus on a fixed spacetime point and observe how fluid elements pass through it.). Such role is played by the spatial coordinates $\sigma^I$ ($I=\bar{1},\bar{2},\bar{3}$) in the fluid spacetime. The worldline (trajectory) of each fluid element is given by a constant-$\sigma^I$ curve and parameterized by $\sigma^{\bar{0}}$. Although the existence of a fluid breaks part of the diffeomorphism symmetry, the following symmetries remains:
\begin{enumerate}
   \item Time-independent reparameterizations of $\sigma^A$
\begin{align}
\sigma^{\bar{0}}\to\sigma^{\bar{0}},\quad \sigma^I\to \sigma'^I(\sigma^I),\quad (I=\bar{1},\bar{2},\bar{3})\,.
\label{spacediffeo}
\end{align} 
	\item Time-diffeomorphisms of $\sigma^{\bar{0}}$
\begin{align}
\sigma^{\bar{0}}\to \sigma'^{\bar{0}}(\sigma^{\bar{0}},\sigma^I),\quad \sigma^I\to \sigma^I,\quad (I=\bar{1},\bar{2},\bar{3})\,.
\label{timediffeo}
\end{align} 
\end{enumerate}
The first condition \eqref{spacediffeo} implies the invariance under a relabeling of fluid elements, while \eqref{timediffeo} implies the time-reparametrization invariance of the fluid element's worldline. These symmetries define a fluid. 

\paragraph{Local temperature.}
In formulating the HydroEFT, we have assumed an initial thermal density matrix $\rho=\frac{1}{Z}e^{-\beta_0 H}$ with an inverse temperature $\beta_0$. Under a perturbation with a timescale much larger than the local thermalization (relaxation) time, thermodynamics quantities including the temperature can fluctuate locally. To describe the local temperature of a fluid, it is convenient to introduce a scalar field $\tau(\sigma)$ which defines the local proper temperature $T(\sigma)$ as
\begin{align}
T(\sigma)\equiv T_0 \, e^{-\tau(\sigma)}
\end{align}
with a reference temperature set by the initial state $T_0=\beta_0^{-1}$. The local temperature $T_{\text{local}}(\sigma)$ normalized by the temporal coordinate
$\sigma^{\bar{0}}$ is defined as
\begin{align}
T_{\text{local}}(\sigma)=E_r T(\sigma)=E_r T_0 e^{-\tau},\quad E_r\equiv \frac{1}{2} \left(\sqrt{-{\sf h}_{1\bar{0}\bar{0}}}+\sqrt{-{\sf h}_{2\bar{0}\bar{0}}}\right) \,.
\end{align} 
From here and onward, we fix $E_r=e^\tau$ by using the time-diffeomorphisms \eqref{timediffeo} to obtain
\begin{align}
T_{\text{local}}(\sigma)=T_0\,,
\end{align} 
and the local proper temperature becomes
\begin{align}
T(\sigma)=\frac{T_0}{E_r}=\frac{1}{\beta(\sigma)}\,.
\end{align} 
These procedures correspond to defining the local temperature by imposing the imaginary fluid time periodicity $\displaystyle\sigma^{\bar{0}}\sim\sigma^{\bar{0}}+ i \beta_0^{-1}$. This gauge choice does not completely fix \eqref{timediffeo} and the following gauge transformation remains:
\begin{align}
    \sigma^{\bar{0}}\to \sigma^{\bar{0}}+\lambda(\sigma^I)\,.
    \label{tempdiffeo}
\end{align}

\paragraph{Dynamical KMS symmetry.}
If the microscopic system has a $Z_2$ symmetry which includes time reversal, the generating function $e^{iW[g_1,g_2]}$ of a system with an initial state in thermal equilibrium (with an inverse temperature $\beta_0$) satisfies the Kubo-Martin-Schwinger (KMS) condition:
\begin{align}
W[g_1(x),g_2(x)]=W[\tilde{g}_1(x) ,\tilde{g}_2(x)]
\label{KMS}
\end{align} 
with
\begin{align}
\tilde{g}_{1\mu\nu}(x)&=\Theta g_{1\mu\nu}\left(t,\vec{x}\right)\,,\quad\tilde{g}_{2\mu\nu}(x)=\Theta g_{2\mu\nu}\left(t+i\beta_0,\vec{x}\right)\,,
\label{gKMS}
\end{align} 
where $\Theta$ is one of the following $Z_2$ transformations $T/PT/CPT$, where $C$, $P$ and $T$ represents the charge conjugation, parity and time reversal, respectively. The explicit choice of this transformation depends on the microscopic system of interest but at least contains $T$.

One can generalize the KMS condition to accommodate systems that are out-of-equilibrium but in local thermal equilibrium. It was proposed in \cite{Glorioso:2017fpd} (and also in \cite{Sieberer:2015hba} in a different context) that the local equilibrium condition can be implemented as a symmetry at the level of the effective action:
\begin{align}
S_{\text{eff}}[X_1(\sigma),X_2(\sigma);{\sf h}_1,{\sf h}_2]=S_{\text{eff}}[\tilde{X}_1(\sigma),\tilde{X}_2(\sigma);\tilde{{\sf h}}_1,\tilde{{\sf h}}_2]
\label{d_KMS}
\end{align} 
with
\begin{align}
\tilde{{\sf h}}_{1 A B}(\sigma)=\Theta {\sf h}_{1 A B} (\sigma^{\bar{0}}, \sigma^I )\,,\quad\tilde{{\sf h}}_{2 A B}(\sigma)=\Theta {\sf h}_{2 A B} (\sigma^{\bar{0}}+i\beta_0, \sigma^I )\,,
\label{hKMS}
\end{align} 
and
\begin{align}
\tilde{X}_1^\mu(\sigma) & =\Theta X_1^\mu (\sigma^{\bar{0}}, \sigma^I)\,,\quad \tilde{X}_2^\mu(\sigma)  =\Theta X_2^\mu (\sigma^{\bar{0}}+i\beta_0, \sigma^I)-i\beta_0 \delta_0^\mu\,.
\label{XKMS}
\end{align} 
\eqref{gKMS} is implied by \eqref{hKMS} and \eqref{XKMS}.
The symmetry associated with the transformations \eqref{hKMS} and \eqref{XKMS} is called the dynamical KMS symmetry. Note that the KMS symmetry \eqref{KMS} is a symmetry based on the temperature of the initial state in equilibrium, whereas the dynamical KMS symmetry \eqref{d_KMS} is a symmetry based on the periodicity of $\sigma^{\bar{0}}$. Therefore, theories with the dynamical KMS symmetry can take any local equilibrium state as the initial state.

\subsection{Ingredients of HydroEFT and symmetries in the classical limit}

To construct the HydroEFT perturbatively, we need to identify a suitable set of variables with nice transformation properties under the symmetries discussed above. In this subsection, we discuss this procedure in more details. We will only consider the classical limit which is adequate for this paper.

\paragraph{$r$-$a$ basis.} 
It is convenient to use the $r$-$a$ basis for considering the classical limit:
\begin{align}
g_{\mu\nu}&=\frac{1}{2} \left(g_{1 \mu\nu}+g_{2 \mu\nu}\right), & \hbar g_{a \mu\nu}&= g_{1 \mu\nu}-g_{2 \mu\nu}\,,\\
X^\mu&=\frac{1}{2}\left(X_1^\mu+X_2^\mu\right), &\hbar X_a^\mu&= X_1^\mu-X_2^\mu\,,\\
{\sf h}_{AB}&=\frac{1}{2} \left({\sf h}_{1 AB}+{\sf h}_{2 AB}\right), &\hbar {\sf h}_{a AB}&= {\sf h}_{1 AB}-{\sf h}_{2 AB}\,.
\end{align} 
As explained in Sec.~\ref{semiclassical}, the spacetime diffeomorphisms in the classical limit are given by \eqref{physicaldiffeo} and \eqref{noisediffeo}. In what follows, we introduce the transformation properties of the fluid diffeomorphisms in the classical limit, along with the corresponding ingredients for constructing the hydro EFT. Furthermore, we present the transformation properties of the dynamical KMS symmetries in the same limit.

\paragraph{Fluid spacetime metric.}

Next, we associate the fluid spacetime metric $h_{AB}$ and the physical spacetime metric $g_{\mu\nu}$ in the classical limit:
\begin{align}
{\sf h}_{AB}(\sigma)\equiv \frac{\partial X^\mu}{\partial \sigma^A}\frac{\partial X^\nu}{\partial \sigma^B}g_{\mu\nu}(X)\,.
\end{align} 
Similarly, the ``noise" metrics ${\sf h}_{aAB}$ on the fluid spacetime reads
\begin{align}
{\sf h}_{aAB}(\sigma)&= \frac{\partial X^\mu}{\partial \sigma^A}\frac{\partial X^\nu}{\partial \sigma^B}\cG_{a\mu\nu}(X)\,,\quad \cG_{a\mu\nu}(X)\equiv g_{a\mu\nu}+\nabla_\mu X_{a\nu}+\nabla_\nu X_{a\mu}\,.
\end{align} 
Notice that $\cG_{a\mu\nu}$ is invariant under the noise diffeomorphism \eqref{noisediffeo}, hence it is a suitable quantity for constructing the EFT.

\paragraph{Ingredients for hydro EFT.}

Now we identify ingredients of the hydro effective action. First, we define the canonically normalized four-velocity $u^\mu$ by
\begin{align}
u^\mu=\frac{1}{\sqrt{-{\sf h}_{\bar{0}\bar{0}}}}\frac{\partial X^\mu}{\partial \sigma^{\bar{0}}}\,,
\quad(u_\mu u^\mu = -1)
\,,
\end{align} 
which is invariant under the fluid diffeomorphisms \eqref{spacediffeo} and \eqref{timediffeo}. The transverse projector orthogonal to $u^\mu$ is defined by
\begin{align}
\Delta^{\mu\nu}=g^{\mu\nu}+u^\mu u^\nu\,,
\end{align}
which satisfies $u_\mu \Delta^{\mu\nu} =0$ and $\Delta^{\mu\nu}\Delta_{\nu\gamma}=\delta^\mu_\gamma$. Also 
it is invariant under \eqref{spacediffeo} and \eqref{timediffeo}.
In the classical limit, we have $E_r=\sqrt{-{\sf h}_{\bar{0}\bar{0}}}$, so that the local proper temperature reads
\begin{align}
T(\sigma)=\frac{T_0}{\sqrt{-h_{\bar{0}\bar{0}}}}=\frac{1}{\beta(\sigma)}\,.
\end{align} 
Note that $\beta(\sigma)$ is invariant under \eqref{tempdiffeo}. Thus, the variables that transform nicely under the symmetries and are convenient for constructing the action are
\begin{align}
\cG_{a \mu\nu},\quad u^\mu,\quad \Delta^{\mu\nu},\quad\beta\,,
\label{ingre}
\end{align} 
and their derivatives.

\paragraph{Dynamical KMS transformations.}
Let us also look at the dynamical KMS transformations for the $r$-$a$ variables in the semiclassical limit. If we recover $\hbar$ explicitly, $\beta_0$ becomes $\hbar\beta_0$, so that we obtain
\begin{align}
\tilde{g}_{\mu \nu}(x) & =\Theta g_{\mu \nu}(x)\,, \quad
&\tilde{g}_{a \mu \nu}(x) &=\Theta g_{a \mu \nu}(x)+i\Theta \mathcal{L}_{\beta_0} g_{\mu \nu}(x)\,, \\
\tilde{X}^\mu(\sigma) & =\Theta X^\mu(\sigma)\,, \quad
&\tilde{X}_a^\mu(\sigma) & =\Theta X_a^\mu(\sigma)-i \Theta \beta^\mu(\sigma)+i \beta_0^\mu\,, \label{XaKMS}\\
\tilde{{\sf h}}_{AB}(\sigma) & =\Theta {\sf h}_{AB}(\sigma)\,, \quad
&\tilde{{\sf h}}_{aAB}(\sigma)  &=\Theta {\sf h}_{aAB}(\sigma)+i \Theta \mathcal{L}_\beta {\sf h}_{AB}(\sigma) \,,
\end{align} 
where $\beta^\mu(x)\equiv\beta(x) u^\mu$, $\beta_0^\mu\equiv\beta_0\delta^\mu_0$, and the physical variables transform as
\begin{align}
\tilde{u}^\mu(x)=\Theta u^\mu(x), \quad \tilde{\beta}(x)=\Theta \beta(x),\quad\tilde{\cG}_{a \mu \nu}(x)=\Theta \cG_{a \mu \nu}(x)+i \Theta \mathcal{L}_\beta g_{\mu \nu}(x) \,.\label{GaKMS}
\end{align} 

\subsection{Entropy current associated with the dynamical KMS invariance \label{local_2nd_law}}

If dynamical KMS symmetry is a symmetry of the action, we can construct a quantity similar to a Noether current by using the Lagrangian formulated in the previous subsection. For HydroEFT, this quantity coincides with the usual entropy current defined in thermodynamics. We will review the construction in this subsection. For simplicity, we only consider the action \eqref{eqn:app_hyaction} which includes derivatives up to first order.

In general, dynamical KMS invariant theories satisfy
\begin{align}
\mathcal{L}_{\text {eff }}=\tilde{\mathcal{L}}_{\text {eff }}-i\nabla_\mu V^\mu\,,
\end{align} 
where $\tilde{\mathcal{L}}_{\text {eff }}$ is obtained by taking the KMS transformation and $x\to x'=\Theta x $, with $\Theta=T,PT,CPT$, of the Lagrangian density. Applying the KMS transformation to \eqref{eqn:app_hyaction} and keeping only upto first order in derivatives, $V^\mu$ is given by
\begin{align}
V^\mu=p_0\beta^\mu \,,
\end{align} 
where we have used the thermodynamics identity $\rho_0 + p_0= -\beta\partial_\beta p_0 $.
On the other hand, using the EOM, the KMS transformation of the Lagrangian becomes 
\begin{align}
\tilde{\mathcal{L}}_{\text {eff }} - \mathcal{L}_{\text {eff }}&=i\nabla_\mu
\left(T^{\mu\nu}  \beta_{\nu} \right)\,.
\label{eqn:surface}
\end{align}
where $T_{\mu\nu}\equiv T_{\mu\nu}^{(0)}+T_{\mu\nu}^{(1)}$. From these calculations, we can define something akin to a Noether current based on the dynamical KMS invariance:
\begin{align}\notag
s^\mu&\equiv V^\mu - T^{\mu\nu}  \beta_{\nu} \\
&=p_0\beta^\mu-T^{\mu\nu}  \beta_{\nu}\,.
\label{ent:def}
\end{align} 
This $s^\mu$ coincides with the usual entropy current in hydrodynamics. 
Considering the divergence of this entropy current $s^\mu$, we have
\begin{align}
\nabla_\mu s^\mu=\frac{\beta}{2}\eta \sigma^{\mu\nu}\sigma_{\mu\nu}+\beta \zeta \vartheta^2\geq0\,,
\label{localsec}
\end{align} 
where we have used the constitution relation of the stress tensor upto first order in derivative and the semi-positive definite properties of $\eta$ and $\xi$, i.e. $\zeta \geq 0,\ \eta \geq 0$, from quantum unitarity condition \eqref{consistency_EFT_6}. For a perfect fluid, the entropy current is conserved. Here, we only presented a proof of the second law upto first order correction. A proof to all orders can be found in \cite{Glorioso:2016gsa}. 


\bibliography{dissipation}{}

\providecommand{\href}[2]{#2}\begingroup\raggedright\begin{thebibliography}{10}

\bibitem{Crossley:2015evo}
M.~Crossley, P.~Glorioso and H.~Liu, \emph{{Effective field theory of
  dissipative fluids}},
  \href{https://doi.org/10.1007/JHEP09(2017)095}{\emph{JHEP} {\bfseries 09}
  (2017) 095} [\href{https://arxiv.org/abs/1511.03646}{{\ttfamily
  1511.03646}}].

\bibitem{Glorioso:2016gsa}
P.~Glorioso and H.~Liu, \emph{{The second law of thermodynamics from symmetry
  and unitarity}},  \href{https://arxiv.org/abs/1612.07705}{{\ttfamily
  1612.07705}}.

\bibitem{Glorioso:2017fpd}
P.~Glorioso, M.~Crossley and H.~Liu, \emph{{Effective field theory of
  dissipative fluids (II): classical limit, dynamical KMS symmetry and entropy
  current}}, \href{https://doi.org/10.1007/JHEP09(2017)096}{\emph{JHEP}
  {\bfseries 09} (2017) 096}
  [\href{https://arxiv.org/abs/1701.07817}{{\ttfamily 1701.07817}}].

\bibitem{Liu:2018kfw}
H.~Liu and P.~Glorioso, \emph{{Lectures on non-equilibrium effective field
  theories and fluctuating hydrodynamics}},
  \href{https://doi.org/10.22323/1.305.0008}{\emph{PoS} {\bfseries TASI2017}
  (2018) 008} [\href{https://arxiv.org/abs/1805.09331}{{\ttfamily
  1805.09331}}].

\bibitem{Sieberer:2015hba}
L.M.~Sieberer, A.~Chiocchetta, A.~Gambassi, U.C.~T\"auber and S.~Diehl,
  \emph{{Thermodynamic Equilibrium as a Symmetry of the Schwinger-Keldysh
  Action}}, \href{https://doi.org/10.1103/PhysRevB.92.134307}{\emph{Phys. Rev.
  B} {\bfseries 92} (2015) 134307}
  [\href{https://arxiv.org/abs/1505.00912}{{\ttfamily 1505.00912}}].

\bibitem{Haehl:2015pja}
F.M.~Haehl, R.~Loganayagam and M.~Rangamani, \emph{{Adiabatic hydrodynamics:
  The eightfold way to dissipation}},
  \href{https://doi.org/10.1007/JHEP05(2015)060}{\emph{JHEP} {\bfseries 05}
  (2015) 060} [\href{https://arxiv.org/abs/1502.00636}{{\ttfamily
  1502.00636}}].

\bibitem{Jensen:2017kzi}
K.~Jensen, N.~Pinzani-Fokeeva and A.~Yarom, \emph{{Dissipative hydrodynamics in
  superspace}}, \href{https://doi.org/10.1007/JHEP09(2018)127}{\emph{JHEP}
  {\bfseries 09} (2018) 127}
  [\href{https://arxiv.org/abs/1701.07436}{{\ttfamily 1701.07436}}].

\bibitem{Jensen:2018hhx}
K.~Jensen, R.~Marjieh, N.~Pinzani-Fokeeva and A.~Yarom, \emph{{An entropy
  current in superspace}},
  \href{https://doi.org/10.1007/JHEP01(2019)061}{\emph{JHEP} {\bfseries 01}
  (2019) 061} [\href{https://arxiv.org/abs/1803.07070}{{\ttfamily
  1803.07070}}].

\bibitem{Minami:2015uzo}
Y.~Minami and Y.~Hidaka, \emph{{Spontaneous symmetry breaking and
  Nambu-Goldstone modes in dissipative systems}},
  \href{https://doi.org/10.1103/PhysRevE.97.012130}{\emph{Phys. Rev. E}
  {\bfseries 97} (2018) 012130}
  [\href{https://arxiv.org/abs/1509.05042}{{\ttfamily 1509.05042}}].

\bibitem{Hongo:2018ant}
M.~Hongo, S.~Kim, T.~Noumi and A.~Ota, \emph{{Effective field theory of
  time-translational symmetry breaking in nonequilibrium open system}},
  \href{https://doi.org/10.1007/JHEP02(2019)131}{\emph{JHEP} {\bfseries 02}
  (2019) 131} [\href{https://arxiv.org/abs/1805.06240}{{\ttfamily
  1805.06240}}].

\bibitem{Hayata:2018qgt}
T.~Hayata and Y.~Hidaka, \emph{{Diffusive Nambu-Goldstone modes in quantum
  time-crystals}},  \href{https://arxiv.org/abs/1808.07636}{{\ttfamily
  1808.07636}}.

\bibitem{Hongo:2019qhi}
M.~Hongo, S.~Kim, T.~Noumi and A.~Ota, \emph{{Effective Lagrangian for
  Nambu-Goldstone modes in nonequilibrium open systems}},
  \href{https://doi.org/10.1103/PhysRevD.103.056020}{\emph{Phys. Rev. D}
  {\bfseries 103} (2021) 056020}
  [\href{https://arxiv.org/abs/1907.08609}{{\ttfamily 1907.08609}}].

\bibitem{Landry:2019iel}
M.J.~Landry, \emph{{The coset construction for non-equilibrium systems}},
  \href{https://doi.org/10.1007/JHEP07(2020)200}{\emph{JHEP} {\bfseries 07}
  (2020) 200} [\href{https://arxiv.org/abs/1912.12301}{{\ttfamily
  1912.12301}}].

\bibitem{Landry:2020ire}
M.J.~Landry, \emph{{Non-equilibrium effective field theory and second sound}},
  \href{https://doi.org/10.1007/JHEP04(2021)213}{\emph{JHEP} {\bfseries 04}
  (2021) 213} [\href{https://arxiv.org/abs/2008.11725}{{\ttfamily
  2008.11725}}].

\bibitem{Landry:2020tbh}
M.J.~Landry, \emph{{Dynamical chemistry: non-equilibrium effective actions for
  reactive fluids}}, \href{https://doi.org/10.1088/1742-5468/ac7a27}{\emph{J.
  Stat. Mech.} {\bfseries 2207} (2022) 073205}
  [\href{https://arxiv.org/abs/2006.13220}{{\ttfamily 2006.13220}}].

\bibitem{Landry:2021kko}
M.J.~Landry, \emph{{Higher-form and (non-)St\"uckelberg symmetries in
  non-equilibrium systems}},
  \href{https://arxiv.org/abs/2101.02210}{{\ttfamily 2101.02210}}.

\bibitem{Fujii:2021nwp}
K.~Fujii and M.~Hongo, \emph{{Effective field theory of fluctuating wall in
  open systems: from a kink in Josephson junction to general domain wall}},
  \href{https://doi.org/10.21468/SciPostPhys.12.5.160}{\emph{SciPost Phys.}
  {\bfseries 12} (2022) 160}
  [\href{https://arxiv.org/abs/2109.10335}{{\ttfamily 2109.10335}}].

\bibitem{Polonyi:2019kjn}
J.~Polonyi, \emph{{Equilibrium properties and decoherence of an open harmonic
  oscillator}}, \href{https://doi.org/10.1088/1751-8121/ab8d08}{\emph{J. Phys.
  A} {\bfseries 53} (2020) 235301}
  [\href{https://arxiv.org/abs/1904.08706}{{\ttfamily 1904.08706}}].

\bibitem{Ota:2024mps}
A.~Ota, \emph{{Fluctuation-dissipation relation in cosmic microwave
  background}},
  \href{https://doi.org/10.1088/1475-7516/2024/05/062}{\emph{JCAP} {\bfseries
  05} (2024) 062} [\href{https://arxiv.org/abs/2402.07623}{{\ttfamily
  2402.07623}}].

\bibitem{Salcedo:2024smn}
S.A.~Salcedo, T.~Colas and E.~Pajer, \emph{{The open effective field theory of
  inflation}}, \href{https://doi.org/10.1007/JHEP10(2024)248}{\emph{JHEP}
  {\bfseries 10} (2024) 248}
  [\href{https://arxiv.org/abs/2404.15416}{{\ttfamily 2404.15416}}].

\bibitem{Hongo:2024brb}
M.~Hongo, N.~Sogabe, M.A.~Stephanov and H.-U.~Yee, \emph{{Schwinger-Keldysh
  effective action for hydrodynamics with approximate symmetries}},
  \href{https://arxiv.org/abs/2411.08016}{{\ttfamily 2411.08016}}.

\bibitem{Salcedo:2024nex}
S.A.~Salcedo, T.~Colas and E.~Pajer, \emph{{An Open Effective Field Theory for
  light in a medium}},  \href{https://arxiv.org/abs/2412.12299}{{\ttfamily
  2412.12299}}.

\bibitem{Akyuz:2023lsm}
C.O.~Akyuz, G.~Goon and R.~Penco, \emph{{The Schwinger-Keldysh coset
  construction}}, \href{https://doi.org/10.1007/JHEP06(2024)004}{\emph{JHEP}
  {\bfseries 06} (2024) 004}
  [\href{https://arxiv.org/abs/2306.17232}{{\ttfamily 2306.17232}}].

\bibitem{Sieberer:2015svu}
L.M.~Sieberer, M.~Buchhold and S.~Diehl, \emph{{Keldysh Field Theory for Driven
  Open Quantum Systems}},
  \href{https://doi.org/10.1088/0034-4885/79/9/096001}{\emph{Rept. Prog. Phys.}
  {\bfseries 79} (2016) 096001}
  [\href{https://arxiv.org/abs/1512.00637}{{\ttfamily 1512.00637}}].

\bibitem{Chen-Lin:2018kfl}
X.~Chen-Lin, L.V.~Delacr\'etaz and S.A.~Hartnoll, \emph{{Theory of diffusive
  fluctuations}},
  \href{https://doi.org/10.1103/PhysRevLett.122.091602}{\emph{Phys. Rev. Lett.}
  {\bfseries 122} (2019) 091602}
  [\href{https://arxiv.org/abs/1811.12540}{{\ttfamily 1811.12540}}].

\bibitem{Chao:2020kcf}
J.~Chao and T.~Schaefer, \emph{{Multiplicative noise and the diffusion of
  conserved densities}},
  \href{https://doi.org/10.1007/JHEP01(2021)071}{\emph{JHEP} {\bfseries 01}
  (2021) 071} [\href{https://arxiv.org/abs/2008.01269}{{\ttfamily
  2008.01269}}].

\bibitem{Cheung:2007st}
C.~Cheung, P.~Creminelli, A.L.~Fitzpatrick, J.~Kaplan and L.~Senatore,
  \emph{{The Effective Field Theory of Inflation}},
  \href{https://doi.org/10.1088/1126-6708/2008/03/014}{\emph{JHEP} {\bfseries
  03} (2008) 014} [\href{https://arxiv.org/abs/0709.0293}{{\ttfamily
  0709.0293}}].

\bibitem{LopezNacir:2011kk}
D.~Lopez~Nacir, R.A.~Porto, L.~Senatore and M.~Zaldarriaga, \emph{{Dissipative
  effects in the Effective Field Theory of Inflation}},
  \href{https://doi.org/10.1007/JHEP01(2012)075}{\emph{JHEP} {\bfseries 01}
  (2012) 075} [\href{https://arxiv.org/abs/1109.4192}{{\ttfamily 1109.4192}}].

\bibitem{Kamenev}
A.~Kamenev, \emph{{Field Theory of Non-Equilibrium Systems}}, Cambridge
  University Press (2011).

\bibitem{BenTov:2021jsf}
Y.~BenTov, \emph{{Schwinger-Keldysh path integral for the quantum harmonic
  oscillator}},  \href{https://arxiv.org/abs/2102.05029}{{\ttfamily
  2102.05029}}.

\bibitem{Weinberg:2008hq}
S.~Weinberg, \emph{{Effective Field Theory for Inflation}},
  \href{https://doi.org/10.1103/PhysRevD.77.123541}{\emph{Phys. Rev. D}
  {\bfseries 77} (2008) 123541}
  [\href{https://arxiv.org/abs/0804.4291}{{\ttfamily 0804.4291}}].

\bibitem{Bellac:2011kqa}
M.L.~Bellac, \emph{{Thermal Field Theory}}, Cambridge Monographs on
  Mathematical Physics, Cambridge University Press (3, 2011),
  \href{https://doi.org/10.1017/CBO9780511721700}{10.1017/CBO9780511721700}.

\bibitem{Bekenstein:1972tm}
J.D.~Bekenstein, \emph{{Black holes and the second law}},
  \href{https://doi.org/10.1007/BF02757029}{\emph{Lett. Nuovo Cim.} {\bfseries
  4} (1972) 737}.

\bibitem{Bekenstein:1973ur}
J.D.~Bekenstein, \emph{{Black holes and entropy}},
  \href{https://doi.org/10.1103/PhysRevD.7.2333}{\emph{Phys. Rev. D} {\bfseries
  7} (1973) 2333}.

\bibitem{Bekenstein:1974ax}
J.D.~Bekenstein, \emph{{Generalized second law of thermodynamics in black hole
  physics}}, \href{https://doi.org/10.1103/PhysRevD.9.3292}{\emph{Phys. Rev. D}
  {\bfseries 9} (1974) 3292}.

\bibitem{Hayward:1993wb}
S.A.~Hayward, \emph{{General laws of black hole dynamics}},
  \href{https://doi.org/10.1103/PhysRevD.49.6467}{\emph{Phys. Rev. D}
  {\bfseries 49} (1994) 6467}.

\bibitem{Hayward:1997jp}
S.A.~Hayward, \emph{{Unified first law of black hole dynamics and relativistic
  thermodynamics}},
  \href{https://doi.org/10.1088/0264-9381/15/10/017}{\emph{Class. Quant. Grav.}
  {\bfseries 15} (1998) 3147}
  [\href{https://arxiv.org/abs/gr-qc/9710089}{{\ttfamily gr-qc/9710089}}].

\bibitem{Ashtekar:2002ag}
A.~Ashtekar and B.~Krishnan, \emph{{Dynamical horizons: Energy, angular
  momentum, fluxes and balance laws}},
  \href{https://doi.org/10.1103/PhysRevLett.89.261101}{\emph{Phys. Rev. Lett.}
  {\bfseries 89} (2002) 261101}
  [\href{https://arxiv.org/abs/gr-qc/0207080}{{\ttfamily gr-qc/0207080}}].

\bibitem{Nielsen:2005af}
A.B.~Nielsen and M.~Visser, \emph{{Production and decay of evolving horizons}},
  \href{https://doi.org/10.1088/0264-9381/23/14/006}{\emph{Class. Quant. Grav.}
  {\bfseries 23} (2006) 4637}
  [\href{https://arxiv.org/abs/gr-qc/0510083}{{\ttfamily gr-qc/0510083}}].

\bibitem{Nielsen:2008cr}
A.B.~Nielsen, \emph{{Black holes and black hole thermodynamics without event
  horizons}}, \href{https://doi.org/10.1007/s10714-008-0739-9}{\emph{Gen. Rel.
  Grav.} {\bfseries 41} (2009) 1539}
  [\href{https://arxiv.org/abs/0809.3850}{{\ttfamily 0809.3850}}].

\bibitem{Debnath:2012fqp}
U.~Debnath, M.~Jamil, R.~Myrzakulov and M.~Akbar, \emph{{Thermodynamics of
  Evolving Lorentzian Wormhole at Apparent and Event Horizons}},
  \href{https://doi.org/10.1007/s10773-014-2159-9}{\emph{Int. J. Theor. Phys.}
  {\bfseries 53} (2014) 4083}
  [\href{https://arxiv.org/abs/1202.1706}{{\ttfamily 1202.1706}}].

\bibitem{Hollands:2024vbe}
S.~Hollands, R.M.~Wald and V.G.~Zhang, \emph{{Entropy of dynamical black
  holes}}, \href{https://doi.org/10.1103/PhysRevD.110.024070}{\emph{Phys. Rev.
  D} {\bfseries 110} (2024) 024070}
  [\href{https://arxiv.org/abs/2402.00818}{{\ttfamily 2402.00818}}].

\bibitem{Visser:2024pwz}
M.R.~Visser and Z.~Yan, \emph{{Properties of dynamical black hole entropy}},
  \href{https://doi.org/10.1007/JHEP10(2024)029}{\emph{JHEP} {\bfseries 10}
  (2024) 029} [\href{https://arxiv.org/abs/2403.07140}{{\ttfamily
  2403.07140}}].

\bibitem{Takeda:2024qbq}
D.~Takeda, \emph{{Coarse-graining black holes out of equilibrium with boundary
  observables on time slice}},
  \href{https://doi.org/10.1007/JHEP05(2024)319}{\emph{JHEP} {\bfseries 05}
  (2024) 319} [\href{https://arxiv.org/abs/2403.07275}{{\ttfamily
  2403.07275}}].

\bibitem{Kong:2024sqc}
D.~Kong, Y.~Tian, H.~Zhang and J.~Zhao, \emph{{Dynamical black hole entropy
  beyond general relativity from the Einstein frame}},
  \href{https://arxiv.org/abs/2412.00647}{{\ttfamily 2412.00647}}.

\bibitem{Nielsen:2007ac}
A.B.~Nielsen and J.H.~Yoon, \emph{{Dynamical surface gravity}},
  \href{https://doi.org/10.1088/0264-9381/25/8/085010}{\emph{Class. Quant.
  Grav.} {\bfseries 25} (2008) 085010}
  [\href{https://arxiv.org/abs/0711.1445}{{\ttfamily 0711.1445}}].

\bibitem{Hayward:2008jq}
S.A.~Hayward, R.~Di~Criscienzo, L.~Vanzo, M.~Nadalini and S.~Zerbini,
  \emph{{Local Hawking temperature for dynamical black holes}},
  \href{https://doi.org/10.1088/0264-9381/26/6/062001}{\emph{Class. Quant.
  Grav.} {\bfseries 26} (2009) 062001}
  [\href{https://arxiv.org/abs/0806.0014}{{\ttfamily 0806.0014}}].

\bibitem{PhysRev.136.B571}
C.W.~Misner and D.H.~Sharp, \emph{Relativistic equations for adiabatic,
  spherically symmetric gravitational collapse},
  \href{https://doi.org/10.1103/PhysRev.136.B571}{\emph{Phys. Rev.} {\bfseries
  136} (1964) B571}.

\bibitem{Kodama:1979vn}
H.~Kodama, \emph{{Conserved Energy Flux for the Spherically Symmetric System
  and the Back Reaction Problem in the Black Hole Evaporation}},
  \href{https://doi.org/10.1143/PTP.63.1217}{\emph{Prog. Theor. Phys.}
  {\bfseries 63} (1980) 1217}.

\bibitem{Murk:2020wkm}
S.~Murk and D.R.~Terno, \emph{{Universal properties of the near-horizon
  geometry}}, \href{https://doi.org/10.1103/PhysRevD.103.064082}{\emph{Phys.
  Rev. D} {\bfseries 103} (2021) 064082}
  [\href{https://arxiv.org/abs/2010.03784}{{\ttfamily 2010.03784}}].

\bibitem{Mann:2021lif}
R.B.~Mann, S.~Murk and D.R.~Terno, \emph{{Surface gravity and the information
  loss problem}},
  \href{https://doi.org/10.1103/PhysRevD.105.124032}{\emph{Phys. Rev. D}
  {\bfseries 105} (2022) 124032}
  [\href{https://arxiv.org/abs/2109.13939}{{\ttfamily 2109.13939}}].

\bibitem{Murk:2021cla}
S.~Murk and D.R.~Terno, \emph{{Physical black holes in semiclassical gravity}},
   in \emph{{16th Marcel Grossmann Meeting on~Recent Developments in
  Theoretical and Experimental General Relativity, Astrophysics and
  Relativistic Field Theories}}, 10, 2021,
  \href{https://doi.org/10.1142/9789811269776_0095}{DOI}
  [\href{https://arxiv.org/abs/2110.12761}{{\ttfamily 2110.12761}}].

\bibitem{Kurpicz:2021kgf}
F.~Kurpicz, N.~Pinamonti and R.~Verch, \emph{{Temperature and
  entropy\textendash{}area relation of quantum matter near spherically
  symmetric outer trapping horizons}},
  \href{https://doi.org/10.1007/s11005-021-01445-7}{\emph{Lett. Math. Phys.}
  {\bfseries 111} (2021) 110}
  [\href{https://arxiv.org/abs/2102.11547}{{\ttfamily 2102.11547}}].

\bibitem{Murk:2023vdw}
S.~Murk and I.~Soranidis, \emph{{Kinematic and energy properties of dynamical
  regular black holes}},
  \href{https://doi.org/10.1103/PhysRevD.108.124007}{\emph{Phys. Rev. D}
  {\bfseries 108} (2023) 124007}
  [\href{https://arxiv.org/abs/2309.06002}{{\ttfamily 2309.06002}}].

\bibitem{Visser:1992qh}
M.~Visser, \emph{{Dirty black holes: Thermodynamics and horizon structure}},
  \href{https://doi.org/10.1103/PhysRevD.46.2445}{\emph{Phys. Rev. D}
  {\bfseries 46} (1992) 2445}
  [\href{https://arxiv.org/abs/hep-th/9203057}{{\ttfamily hep-th/9203057}}].

\bibitem{Abreu:2010ru}
G.~Abreu and M.~Visser, \emph{{Kodama time: Geometrically preferred foliations
  of spherically symmetric spacetimes}},
  \href{https://doi.org/10.1103/PhysRevD.82.044027}{\emph{Phys. Rev. D}
  {\bfseries 82} (2010) 044027}
  [\href{https://arxiv.org/abs/1004.1456}{{\ttfamily 1004.1456}}].

\bibitem{Pielahn:2011ra}
M.~Pielahn, G.~Kunstatter and A.B.~Nielsen, \emph{{Dynamical Surface Gravity in
  Spherically Symmetric Black Hole Formation}},
  \href{https://doi.org/10.1103/PhysRevD.84.104008}{\emph{Phys. Rev. D}
  {\bfseries 84} (2011) 104008}
  [\href{https://arxiv.org/abs/1103.0750}{{\ttfamily 1103.0750}}].

\bibitem{Cropp:2013zxi}
B.~Cropp, S.~Liberati and M.~Visser, \emph{{Surface gravities for non-Killing
  horizons}},
  \href{https://doi.org/10.1088/0264-9381/30/12/125001}{\emph{Class. Quant.
  Grav.} {\bfseries 30} (2013) 125001}
  [\href{https://arxiv.org/abs/1302.2383}{{\ttfamily 1302.2383}}].

\bibitem{Hawking:1975vcx}
S.W.~Hawking, \emph{{Particle Creation by Black Holes}},
  \href{https://doi.org/10.1007/BF02345020}{\emph{Commun. Math. Phys.}
  {\bfseries 43} (1975) 199}.

\bibitem{Baggioli:2020haa}
M.~Baggioli and M.~Landry, \emph{{Effective Field Theory for Quasicrystals and
  Phasons Dynamics}},
  \href{https://doi.org/10.21468/SciPostPhys.9.5.062}{\emph{SciPost Phys.}
  {\bfseries 9} (2020) 062} [\href{https://arxiv.org/abs/2008.05339}{{\ttfamily
  2008.05339}}].

\end{thebibliography}\endgroup
\bibliographystyle{JHEP}

\end{document}